\documentclass[aps,twocolumn,nofootinbib,groupedaddress,superscriptaddress,longbibliography]{revtex4-1}

\usepackage{amsmath,amssymb}
\usepackage{graphicx}
\usepackage{multirow}
\usepackage{bm}
\usepackage{mathtools}
\usepackage{dsfont}
\usepackage{amsfonts}
\usepackage{xcolor}
\usepackage{ulem}
\usepackage{url}

\usepackage[colorlinks=true,linkcolor=magenta,citecolor=magenta,hypertexnames=false]{hyperref}

\def\ba#1\ea{\begin{align}#1\end{align}}
\def\bg#1\eg{\begin{gather}#1\end{gather}}
\def\bpm{\begin{pmatrix}}
\def\epm{\end{pmatrix}}
\newcommand{\nn}{\nonumber}

\newcommand{\bb}[1]{{\boldsymbol #1}}
\newcommand{\bx}{\bb x}
\newcommand{\bk}{\bb k}
\newcommand{\bR}{\bb R}

\newcommand{\mc}[1]{\mathcal{#1}}

\newcommand{\dg}{\dagger}

\newcommand{\sg}{\sigma}
\newcommand{\vph}{\varphi}

\newcommand{\ket}[1]{|#1\rangle}
\newcommand{\bra}[1]{\langle#1|}
\newcommand{\brk}[2]{\langle#1|#2\rangle}
\newcommand{\kbr}[2]{|#1\rangle\langle#2|}

\newcommand{\Rf}[1]{Ref.~\cite{#1}}

\newcommand{\eq}[1]{Eq.~\eqref{#1}}
\newcommand{\eqs}[1]{Eqs.~\eqref{#1}}
\newcommand{\fig}[1]{Fig.~\ref{#1}}
\newcommand{\figs}[1]{Figs.~\ref{#1}}

\newcommand{\bsk}[1]{\ket{\vph_{#1}(\bk)}}
\newcommand{\bsb}[1]{\bra{\vph_{#1}(\bk)}}
\newcommand{\bs}[2]{\ket{\vph_{#1}(#2)}}
\newcommand{\bskL}[3]{\ket{\vph#1_{#2}(#3)}}
\newcommand{\bsbL}[3]{\bra{\vph#1_{#2}(#3)}}

\newcommand{\cm}[1]{\overline{#1}}

\newcommand{\symeigv}{\xi_\sg}
\newcommand{\bchi}{\bb \chi}

\newcommand{\uef}[1]{\ket{\hat{u}_{#1}(\bk)}}
\newcommand{\uefL}[2]{\ket{\hat{u}#1_{#2}(\bk)}}

\newcommand{\ourtitle}{
General construction of flat bands with and without band crossings based on wave function singularity}

\allowdisplaybreaks

\begin{document}
\title{\textbf{\ourtitle}}

\author{Yoonseok \surname{Hwang}}
\affiliation{Center for Correlated Electron Systems, Institute for Basic Science (IBS), Seoul 08826, Korea}
\affiliation{Department of Physics and Astronomy, Seoul National University, Seoul 08826, Korea}
\affiliation{Center for Theoretical Physics (CTP), Seoul National University, Seoul 08826, Korea}

\author{Jun-Won \surname{Rhim}}
\affiliation{Center for Correlated Electron Systems, Institute for Basic Science (IBS), Seoul 08826, Korea}
\affiliation{Department of Physics and Astronomy, Seoul National University, Seoul 08826, Korea}
\affiliation{Department of Physics, Ajou University, Suwon 16499, Korea}

\author{Bohm-Jung \surname{Yang}}
\email{bjyang@snu.ac.kr}
\affiliation{Center for Correlated Electron Systems, Institute for Basic Science (IBS), Seoul 08826, Korea}
\affiliation{Department of Physics and Astronomy, Seoul National University, Seoul 08826, Korea}
\affiliation{Center for Theoretical Physics (CTP), Seoul National University, Seoul 08826, Korea}

\begin{abstract}
In this work, we develop a systematic method of constructing flat-band models with and without band crossings.
Our construction scheme utilizes the symmetry and spatial shape of a compact localized state (CLS) and also the singularity of the flat-band wave function obtained by a Fourier transform of the CLS (FT-CLS).
In order to construct a flat-band model systematically using these ingredients, we first choose a CLS with a specific symmetry representation in a given lattice.
Then, the singularity of FT-CLS indicates whether the resulting flat band exhibits a band crossing point or not.
A tight-binding Hamiltonian with the flat band corresponding to the FT-CLS is obtained by introducing a set of basis molecular orbitals, which are orthogonal to the FT-CLS.
Our construction scheme can be systematically applied to any lattice so that it provides a powerful theoretical framework to study exotic properties of both gapped and gapless flat bands arising from their wave function singularities.
\end{abstract}

\maketitle

\section{Introduction \label{sec:Intro}}
In past decades, the studies of flat-band (FB) systems~\cite{lieb1989two,aoki1996hofstadter,huber2010bose,weeks2012flat,julku2016geometric,ramachandran2017chiral,misumi2017new,pal2018flat,mizoguchi2019flat,hwang2020geometric,kuno2020flat,lin2020chiral,morfonios2021flat,peri2021fragile,regnault2021catalogue} have been mostly focused on strong correlation physics such as fractional quantum Hall effect~\cite{regnault2011fractional,tang2011high,sun2011nearly,neupert2011fractional,andrews2020fractional}, ferromagnetism~\cite{mielke1991ferromagnetic,mielke1991ferromagnetism,mielke1993ferromagnetism}, Wigner crystallization~\cite{wu2007flat}, and so on, which originate from the flat energy dispersion.
On the other hand, the recent discoveries of FBs in the kagome materials~\cite{ye2018massive,li2018realization,kang2020topological} and twisted bilayer graphene~\cite{bistritzer2011moire,cao2018correlated,cao2018unconventional} have demonstrated that the nontrivial topological and geometric properties can also arise in FB systems due to the characteristics of the FB wave functions.
For example, in kagome materials, the nearly flat bands have been attracted great attention due to their nontrivial band topology with nonzero Chern number and time-reversal $\mathbb{Z}_2$ invariant.
Similarly, the nearly flat bands in twisted bilayer graphene have fragile band topology~\cite{po2018fragile,song2019all,po2019faithful,ahn2019failure}.
Such nearly flat bands with nontrivial topology can enhance the superfluity weight~\cite{peotta2015superfluidity,xie2020topology}, related to their nontrivial wave-function geometry.

Moreover, it has been recently pointed out that FBs with band crossing points can be an ideal platform for studying new types of topological and geometric properties
related to the singularity of FB wave functions.
A FB can be classified according to the presence or absence of singular points in its wave function.
A FB with singular band crossing points is called a singular FB (SFB)~\cite{rhim2019classification,rhim2021singular}.
Otherwise, the FB is classified as a nonsingular FB (NSFB).
The band crossing points of SFBs are enforced by symmetry representation (SR)~\cite{hwang2021flat} of compact localized state (CLS)~\cite{sutherland1986localization,vidal1998aharonov,vidal2001disorder,mukherjee2015observation,read2017compactly,rontgen2018compact,ma2020direct,yang2021symmetry}, which is a characteristic eigenstate of a FB strictly confined within a finite region in real space.
Interestingly, recent studies have shown that the singularity of the SFB gives anomalous Landau level spectrum which manifests the maximum quantum distance of the FB wave functions~\cite{rhim2020quantum,hwang2021wave}.
Also, the degeneracy lifting at the singular band crossing points can induce nearly flat bands with nontrivial topological properties~\cite{rhim2019classification,hwang2021flat,ma2020spin}.

For systematic investigation of fundamental physical properties of FB systems, a general method of constructing FB models plays a quintessential role.
Several general schemes for constructing FB models have been proposed recently.
In Refs.~\cite{maimaiti2017compact,maimaiti2019universal,maimaiti2021flat}, FB models are constructed by using the flat band generator in one and two dimensions.
On the other hand, FB models in several lattices including Kagome and pyrochlore lattices are constructed using the molecular-orbital representation~\cite{mizoguchi2019molecular,mizoguchi2020systematic,mizoguchi2020type,mizoguchi2021flat}.
Also, general construction schemes~\cite{ma2020spin,chiu2020fragile,cualuguaru2021general} for obtaining FBs with and without band crossing points have been proposed on the basis of graph theories~\cite{mielke1991ferromagnetic,mielke1991ferromagnetism,mielke1993ferromagnetism,kollar2019line}, symmetry indicators~\cite{kruthoff2017topological,po2017symmetry,watanabe2018structure}, and topological quantum chemistry~\cite{bradlyn2017topological,cano2018building,elcoro2020magnetic}.
These schemes are applied to bipartite lattices, where two mutually disjoint sets of sublattices have unequal cardinalities, as well as split and line graph lattices.
However, a general construction scheme utilizing the singularity of FB wave functions does not exist yet, while the singularity turned out to be crucial in understanding physical properties of FBs.
%

\begin{figure}[t!]
\centering
\includegraphics[width=0.48\textwidth]{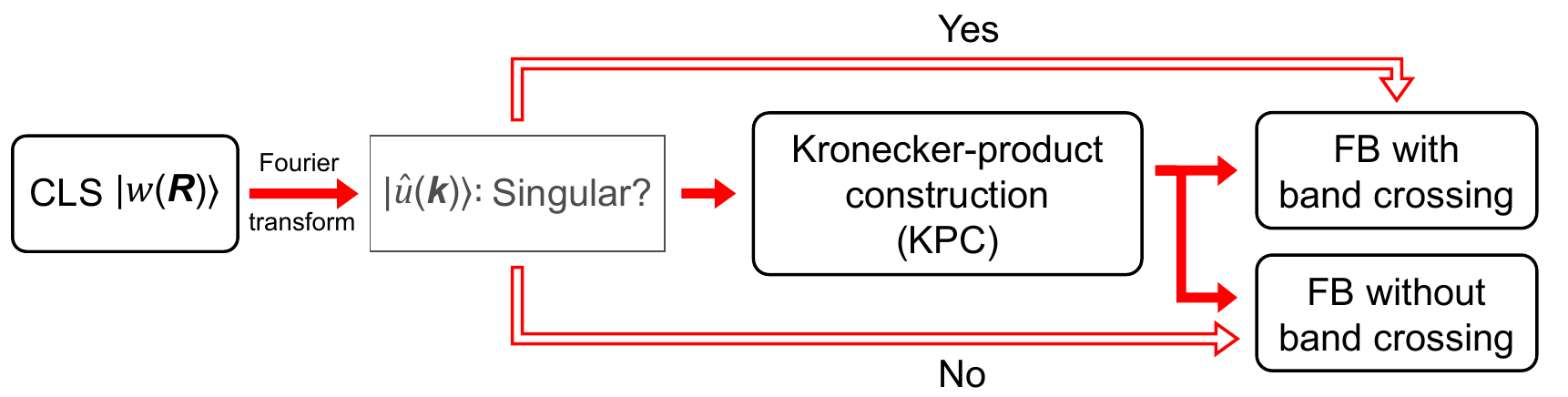}
\caption{
Construction scheme in this work.
First, consider a compact localized state (CLS) in a given lattice.
Then, determine whether the corresponding Fourier transform of CLS (FT-CLS) $\uef{}$ is singular or not.
When the FT-CLS is nonsingular, then we obtain a flat band (FB) without a band crossing.
On the other hand, when $\uef{}$ has singular points, then the resulting FB model exhibits a FB with band crossing points at the singular points.
For a given FT-CLS, the FB tight-binding Hamiltonian can be obtained by choosing basis molecular orbitals which are orthogonal to the FT-CLS, appropriately.
}
\label{Fig1}
\end{figure}

In this work, we propose a systematic scheme for constructing FB models with and without band crossing points, mostly focusing on nondegenerate FBs.
Figure~\ref{Fig1} shows the structure of our construction scheme, which is based on SR under unitary symmetry of a CLS.
A CLS transforms like Wannier function under symmetry with a specific SR.
Then, we obtain a Fourier transform of CLS (FT-CLS) $\uef{}$.
Importantly, the presence or absence of singular points in FT-CLS $\uef{}$ determines whether the resulting FB model has band crossings or not~\cite{hwang2021flat}, even before we construct a specific tight-binding Hamiltonian.
To construct a FB model having $\uef{}$ as a FT-CLS, we introduce basis molecular orbitals (BMOs) orthogonal to $\uef{}$,
which enable us to find a relevant hopping structure.
Then the tight-binding Hamiltonian with the FB is given by summing the Kronecker products of BMOs.
We apply our construction scheme to various lattice systems and illustrate how FB models with and without band crossing points can be systematically constructed.

The rest of this paper is organized as follows. 
First, we introduce our construction scheme by focusing on the Lieb lattice and constructing NSFBs and SFBs in Sec.~\ref{sec:Lieb}.
Then, our construction scheme is further elaborated in Sec.~\ref{sec:KPC} such that it can generally be applied to any lattice system with any symmetry.
In Sec.~\ref{app:examples}, we apply our construction scheme to various lattice systems.
While we focus on nondegenerate FB mainly, we show that degenerate FB can also be obtained through our construction scheme in Sec.~\ref{sec:deg_flat}.
Finally, we summarize and discuss possible extensions of this work in Sec.~\ref{sec:summary}.

\section{Flat band models in the Lieb lattice \label{sec:Lieb}}
We illustrate the general idea of our FB construction scheme focusing on the FB in the Lieb lattice~\cite{lieb1989two}.
More detailed description of the general construction scheme is provided in Sec.~\ref{sec:KPC}.
The unit cell of the Lieb lattice is composed of three sublattice sites as shown in \fig{Fig2}(a).
The orbitals on sublattice sites $\bR+\bx_\alpha$ are described by $\ket{\bR,\alpha}$ ($\alpha=1,2,3$), where $\bR$ denotes a (Bravais) lattice vector.
Considering only the nearest-neighbor hopping $t_0$ [see \fig{Fig2}(b)], we obtain a tight-binding Hamiltonian
\ba
H^{(0)}_{\rm Lieb}(\bk) = t_0
\bpm 0 & 1+Q_1 & 1+Q_2 \\ 1+\cm{Q_1} & 0 & 0 \\ 1+\cm{Q_2} & 0 & 0 \epm
\label{eq:Lieb_Original}
\ea
where
\ba
Q_i=e^{-i \bk \cdot \bb a_i
\label{eq:Q_Lieb}}
\ea
with lattice vectors $\bb a_i$ ($i=1,2$), and $\cm{x}$ denotes the complex conjugation of $x$.

The band structure of $H^{(0)}_{\rm Lieb}(\bk)$ exhibits a FB at zero energy which has a band crossing with three-fold degeneracy at $M=(\pi,\pi)$ [see \fig{Fig2}(c)].
One way of explaining the existence of a FB in the model $H^{(0)}_{\rm Lieb}(\bk)$ is to consider chiral symmetry $C$.
Namely, $H^{(0)}_{\rm Lieb}(\bk)$ is symmetric under chiral symmetry, i.e. $U_C H^{(0)}_{\rm Lieb}(\bk) {U_C}^{-1} = - H^{(0)}_{\rm Lieb}(\bk)$, where $U_C={\rm Diag}(-1,1,1)$ denotes chiral symmetry operator.
In general, in chiral symmetric systems, when the chiral symmetry operator $U_C$ satisfies ${\rm Tr} [U_C]=\pm n_F$,
there must be $n_F$ number of FBs at zero energy~\cite{lieb1989two}.
In the case of $H^{(0)}_{\rm Lieb}(\bk)$, as $n_F=1$, a single FB at zero energy can appear.

However, chiral symmetry is not essential for constructing FB models in general.
In the following, we present various FB models in the Lieb lattice, other than the chiral-symmetric model $H^{(0)}_{\rm Lieb}(\bk)$, by focusing on symmetry and algebraic properties of CLS.

\subsection{Compact localized state and its Fourier transform \label{subsec:CLS_Lieb}}
For a given FB, one can always find a relevant CLS~\cite{read2017compactly}.
In the case of $H^{(0)}_{\rm Lieb}(\bk)$, the normalized eigenstate of the FB $\ket{u_{\rm Lieb}(\bk)}$ is given by $\ket{u_{\rm Lieb}(\bk)} = \mc{N}(\bk) \uef{\rm Lieb}$ where $\mc{N}(\bk)=(4+2\cos \bk \cdot \bb a_1 + 2\cos \bk \cdot \bb a_2)^{-1/2}$ is a normalization factor. For the analysis of FB, we introduce the following form of an unnormalized eigenstate $\uef{\rm Lieb}$, given by
\ba
\uef{\rm Lieb} =(0,-1-Q_2,1+Q_1).
\label{eq:FT-CLS_Lieb}
\ea
Note that $\uef{\rm Lieb}$ is expressed by Laurent polynomials in $Q_{1,2}$, which is guaranteed for any FB models with finite-ranged hoppings~\cite{read2017compactly,rhim2019classification}.
A CLS $\ket{w(\bR)}$, defined for the unit cell at the lattice position $\bR$, is obtained by 
\ba
\ket{w(\bR)} = \sum_{\bR',\alpha} \, S_\alpha(\bR') \, \ket{\bR+\bR',\alpha},
\label{eq:CLS_def1}
\ea
where 
\ba
S_\alpha(\bR) = \frac{1}{N_{\rm cell}} \sum_{\bk} \, e^{i\bk\cdot\bR} \ \uef{}_\alpha,
\label{eq:CLS_def2}
\ea
in which $N_{\rm cell}$ is the number of unit cells.

\begin{figure}[t!]
\centering
\includegraphics[width=0.48\textwidth]{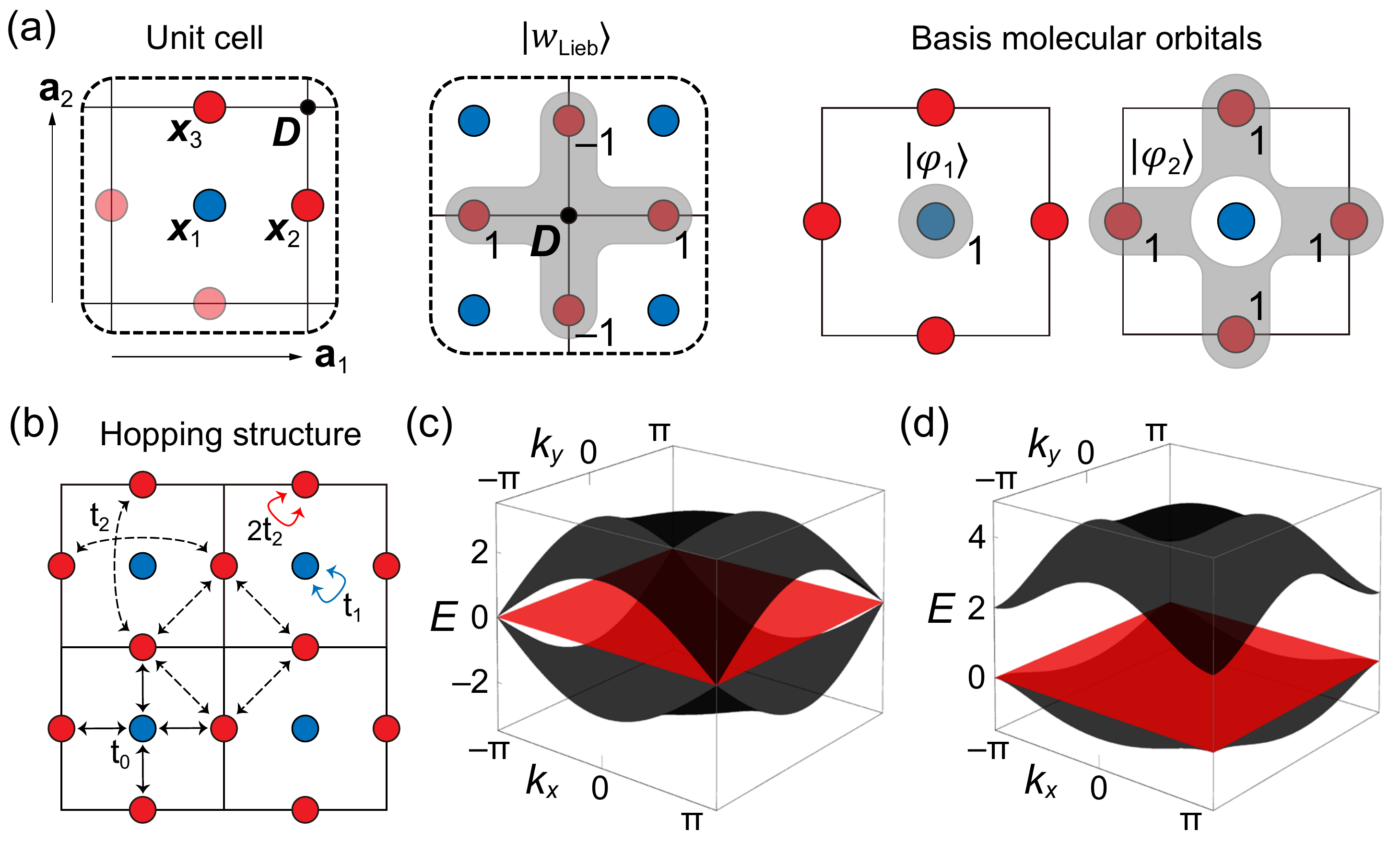}
\caption{
Description of the Lieb lattice model.
(a) The unit cell, compact localized state (CLS), and basis molecular orbitals (BMOs).
A unit cell is composed of three sublattice sites located at $\bx_1=(0,0)$, $\bx_2=(1/2,0)$, and $\bx_3=(0,1/2)$, respectively.
The Wyckoff position $\bb D$ is indicated by the black dot.
The primitive lattice vectors are $\bb a_1=(1,0)$ and $\bb a_2=(0,1)$.
The gray regions indicate the shapes of the CLS $\ket{w_{\rm Lieb}(\bR)}$ and BMOs $\ket{\vph_{1,2}(\bR)}$.
The number near each site denotes the amplitude of the state.
(b) Description of the hopping processes of the model. The same type of arrows
represents the hopping with the same strength.
(c) Band structure of $H^{(0)}_{\rm Lieb}(\bk)$ for $t_0=1.0$.
The FB at zero energy has a band crossing with three-fold degeneracy at $M=(\pi,\pi)$.
(d) Band structure of $H_{\rm Lieb}(\bk)$ for $(t_0,t_1,t_2)=(1.0,2.0,0.2)$.
The FB still has a band crossing at $M$, but only with the lowest band.
}
\label{Fig2}
\end{figure}

For convenience, we refer to the unnormalized eigenstate $\uef{}$ as a Fourier transform of CLS (FT-CLS), considering its relation to the CLS.
When the hoppings are finite ranged, $S_\alpha(\bR')$ is nonzero only inside a finite region.
We refer to such a compact region with nonzero $S_\alpha(\bR')$  as a \textit{shape} of the CLS.
According to \eq{eq:CLS_def1}, the CLS of $H^{(0)}_{\rm Lieb}(\bk)$ is given by $\ket{w_{\rm Lieb}(\bR)} = - \ket{\bR,2} - \ket{\bR+\bb a_2,2} + \ket{\bR,3} + \ket{\bR+\bb a_1,3}$.
Its shape is drawn schematically in \fig{Fig2}(a).
We note that the CLS $\ket{w_{\rm Lieb}(\bR)}$ is obtained by using the known FT-CLS $\uef{\rm Lieb}$ above.
However, when we construct a FB model in general, it is more convenient to choose a CLS first and then find the corresponding FT-CLS, as discussed in detail below.

\subsection{Kronecker-product construction \label{subsec:KPC_Lieb}}
Now, we introduce a general method of systematically constructing a tight-binding Hamiltonian with a FB for a given CLS.
For convenience, we continue to consider the CLS $\ket{w_{\rm Lieb}(\bR)}$ in \fig{Fig2}(a).
Then, our goal is to construct a FB model having the FT-CLS $\uef{\rm Lieb}$ as a FB eigenstate.
Such a procedure generally requires a fine tuning of model parameters.
However, in our construction scheme, we bypass this problem by introducing \textit{basis molecular orbitals} (BMOs).
We note that the BMOs play a similar role as the molecular orbitals used in Refs.~\cite{bilitewski2018disordered,mizoguchi2019molecular,mizoguchi2020systematic,mizoguchi2020type,mizoguchi2021flat}.

We determine the BMOs first in momentum space using the fact that they are orthogonal to the given FT-CLS.
For the FT-CLS $\uef{\rm Lieb}$, we introduce two BMOs $\bsk{1,2}$ satisfying $\brk{\vph_{1,2}(\bk)}{\hat{u}_{\rm Lieb}(\bk)}=0$ as
\ba
\bsk{1}=(1,0,0), \quad \bsk{2}=(0,1+\cm{Q_1},1+\cm{Q_2}).
\label{eq:Lieb_basis}
\ea
Note that the BMOs are not necessary to be mutually orthogonal.
Also, as in the case of FT-CLS, we require that each element of BMOs is a Laurent polynomial to ensure finite-ranged hoppings.
Note that the BMOs are not normalized because of this condition.
Then, the real-space representations of BMOs $\bskL{}{1,2}{\bR}$, illustrated in \fig{Fig2}(a), are obtained by a Fourier transform in a similar way as in \eqs{eq:CLS_def1} and \eqref{eq:CLS_def2}.

With the chosen CLS and BMOs, we construct a FB Hamiltonian on the Lieb lattice as
\ba
H_{\rm Lieb}(\bk) = \sum_{a,b=1}^2 \, f_{ab}(\bk) \, h_{ab}(\bk),
\label{eq:KPC_Lieb}
\ea
where $h_{ab}(\bk)=h_{ba}(\bk)^{*}=\bsk{a}\bsb{b}$ and $f_{ab}(\bk)$ are Laurent polynomials of $Q_{1,2}$.
Note that $f_{ab}(\bk)=\cm{f_{ba}(\bk)}$ to ensure hermiticity of $H_{\rm Lieb}(\bk)$.
This form of the Hamiltonian is guaranteed to have a FB at zero energy because the orthogonality $\brk{\vph_a(\bk)}{\hat{u}_{\rm Lieb}(\bk)}=0$ leads to $H_{\rm Lieb}(\bk) \uef{\rm Lieb}=0$.
Since $H_{\rm Lieb}(\bk)$ is expressed by a sum of Kronecker products of BMOs, we refer to our method for constructing a FB Hamiltonian as \textit{Kronecker-product construction} (KPC) scheme, and we call the resulting Hamiltonian a KPC Hamiltonian.

For simplicity, we choose $f_{11}(\bk)=t_1$, $f_{12}(\bk)=t_0$, and $f_{22}(\bk)=t_2$ where $t_{0,1,2} \in \mathbb{R}$.
The relevant hopping structure is illustrated in \fig{Fig2}(b).
[The hopping parameters between the orbitals on sublattice sites can be obtained by Fourier transform given in \eq{eq:hopping}.]
Note that $H_{\rm Lieb}(\bk)$ reduces to $H^{(0)}_{\rm Lieb}(\bk)$ when $t_1=t_2=0$.
Through the KPC scheme, it is straightforward to include more hoppings ($t_{1,2}$) to the original Lieb model $H^{(0)}_{\rm Lieb}(\bk)$, beyond the nearest-neighbor hopping $t_0$, while keeping a FB.

We comment two important properties of KPC Hamiltonian $H_{\rm Lieb}(\bk)$, which are generalized for arbitrary cases in Sec.~\ref{sec:KPC}.
First, we have two BMOs $\bsk{1,2}$, i.e., $n_B=2$ where $n_B$ denotes the number of BMOs .
While $n_B$ does not change in the construction, the number of independent BMOs $N_B(\bk)$ can vary at each $\bk$ when the BMOs are viewed as complex-valued vectors at each $\bk$.
Specifically, $N_B(\bk)$ is equal to 2 at $\bk$ except at the momentum $M$, while $N_B(\bk)$ decreases to 1 at $\bk=M$ since $\ket{\vph_2(M)}=(0,0,0)$.
The number of flat bands $n_F$ is given by $n_F=n_{\rm tot}-{\rm max} \, N_B(\bk)$ for a general choice of $f_{ab}(\bk)$ where $n_{\rm tot}$ indicates the total number of bands.
Hence, $H_{\rm Lieb}(\bk)$ has a single FB ($n_F=1$).
On the other hand, at the momentum $\bk_*$ where $N_B(\bk_*)$ does not take its maximum value, ${\rm max} \, N_B(\bk)$, the FB has band crossing with dispersive bands at $\bk_*$.
The degree of degeneracy at the band crossing point depends on the detailed value of $f_{ab}(\bk_*)$.
At $M$, $H_{\rm Lieb}(\bk)$ takes the form $f_{11}(M) h_{11}(M)$ and a FB has a band crossing according to $N_B(M)(=1)< {\rm max} \, N_B(\bk)(=2)$.
When $f_{11}(M)$ is nonzero, the degeneracy is two, which is the same to the value of $n_{\rm tot} - N_B(\bk)$.
On the other hand, if $f_{11}(M)$ is zero, $H_{\rm Lieb}(M)$ is the $3 \times 3$ zero matrix and the degeneracy becomes three-fold as in $H^{(0)}_{\rm Lieb}(\bk)$.
We also note that the number of BMOs is not necessary to be equal to ${\rm max} \, N_B(\bk)$ in general.

Second, a KPC Hamiltonian can always be symmetrized such that the Hamiltonian has symmetries that the chosen CLS possesses.
For the details of symmetrization, see Sec.~\ref{sec:KPC} and Appendix~\ref{app:symmetrization}.
For a brief illustration, let us focus on four-fold rotation symmetry $C_4$ in the Lieb lattice.
The CLS $\ket{w_{\rm Lieb}(\bR)}$ and FT-CLS $\uef{\rm Lieb}$ are symmetric under $C_4$:
\bg
\label{eq:SR_CLS_Lieb}
\hat{C_4} \ket{w_{\rm Lieb}(\bR)} = -\ket{w_{\rm Lieb}(O_{C_4}\bR-\bb a_2)}, \\
\label{eq:SR_FT-CLS_Lieb}
U_{C_4}(\bk) \uef{\rm Lieb} = -Q_2 \ket{\hat{u}_{\rm Lieb}(O_{C_4}\bk)},
\eg
where $O_{C_4}$ denotes a $C_4$ rotation in real space, $O_{C_4}\bR = (-R_y,R_x)$, and the symmetry operator for $C_4$ is defined as
\ba
U_{C_4}(\bk) = \bpm 1 & 0 & 0 \\ 0 & 0 & Q_2 \\ 0 & 1 & 0 \epm.
\ea
Then, $H_{\rm Lieb}(\bk)$ is symmetric under $C_4$, as long as $f_{ab}(\bk) = f_{ab}(O_{C_4} \bk)$ is satisfied, i.e.
\ba
H_{\rm Lieb}(O_{C_4} \bk)= U_{C_4}(\bk) H_{\rm Lieb}(\bk) U_{C_4}(\bk)^\dg.
\label{eq:Lieb_C4rel}
\ea
This follows because the chosen BMOs transform trivially under $C_4$: $U_{C_4}(\bk) \bsk{a} = \bs{a}{O_{C_4} \bk}$, hence $U_{C_4}(\bk) h_{ab}(\bk) U_{C_4}(\bk)^\dg = h_{ab}(O_{C_4} \bk)$.

\subsection{Singular and nonsingular flat bands \label{subsec:singular_Lieb}}
The band structure for $(t_0,t_1,t_2)=(1.0,2.0,0.2)$, shown in \fig{Fig2}(d), exhibits a two-fold band crossing between the flat and dispersive bands at $M$.
For nonzero $t_1$, the three-fold degeneracy at $M$ that exists when $t_1=0$ splits into a two-fold degeneracy and a nondegenerate state.
Nevertheless, for any choice of $f_{ab}(\bk)$, one can always find a band crossing at $M$.
In fact, the band crossing of the FB at $M$ is enforced by SR of the CLS~\cite{hwang2021flat}.
To understand this, we first notice that the FT-CLS $\uef{\rm Lieb}$ becomes zero at $\bk=M$, i.e. $\ket{\hat{u}_{\rm Lieb}(M)}=(0,0,0)$.
When a FT-CLS $\uef{}$ becomes zero at some momenta which we call singular points, the corresponding FB is called a singular FB (SFB)~\cite{rhim2019classification,rhim2020quantum,rhim2021singular,hwang2021flat}.
Moreover, a SFB must have band crossings with other dispersive bands at the singular points~\cite{rhim2019classification,hwang2021flat}.
Contrary to the case of the SFB, the FT-CLS of NSFB is nonzero everywhere in the BZ.
Also, a NSFB does not have a band crossing with other bands unless it is fine tuned.

In our construction scheme, as a FT-CLS $\uef{}$ is readily obtained from the shape of CLS, the presence or absence of band crossings, directly related to its singular points, can also be determined from the outset.
In the case of $\ket{w_{\rm Lieb}(\bR)}$, the corresponding FT-CLS $\uef{\rm Lieb}$ is singular at $M$ and the resulting KPC Hamiltonian must exhibit a FB with a band crossing at $M$.

One can also obtain a NSFB in the Lieb lattice by choosing a CLS whose FT-CLS is nonsingular.
For example, we can deform $\ket{w_{\rm Lieb}(\bR)}$ slightly and consider $\ket{w'_{\rm Lieb}(\bR)}
= -m\ket{\bR,2}-\ket{\bR+\bb a_2,2} + m\ket{\bR,3}+\ket{\bR+\bb a_1,3}$ with a real parameter $m$.
The corresponding FT-CLS $\uefL{'}{\rm Lieb}=(0,-m-Q_2,m+Q_1)$ is nonzero everywhere in the BZ unless $m=\pm 1$.
We note that the parameter $m$ breaks $C_4$ and a KPC Hamiltonian without symmetry represents a FB model with a complicated hopping structure.

Recently, a close connection between the singularity of FT-CLS and crystalline symmetries has been pointed out in \Rf{hwang2021flat}.
That is, a FB must be a SFB independent of the detailed shape, when its symmetry representation (SR) satisfies a certain condition.
The relevant band degeneracy point of the FB is called the SR-enforced band crossing points.
Here let us briefly recap the key idea and discuss symmetry property of CLSs in the Lieb lattice.
In a $C_4$-symmetric lattice such as the Lieb lattice, there are three types of maximal Wyckoff positions at $\bb A=(0,0)$, $\bb B=\{(1/2,0),(0,1/2)\}$, and $\bb D=(1/2,1/2)$, respectively.
The three sublattice sites of the Lieb lattice correspond to the Wyckoff positions $\bb A$ and $\bb B$.
The CLS $\ket{w_{\rm Lieb}(\bR)}$ is centered at the Wyckoff position $\bb D$ and has a $C_4$ eigenvalue $-1$, as shown in \fig{Fig2}(a).
Accordingly, the CLS and FT-CLS transform under $C_4$ as \eqs{eq:SR_CLS_Lieb} and \eqref{eq:SR_FT-CLS_Lieb}.
Importantly, \eq{eq:SR_FT-CLS_Lieb} leads to $\ket{\hat{u}_{\rm Lieb}(M)}=(u_1,0,0)$ with $u_1 \in \mathbb{C}$.
When a CLS does not occupy the first sublattice as like $\ket{w_{\rm Lieb}(\bR)}$, $u_1$ is zero identically.
Hence, $\ket{\hat{u}_{\rm Lieb}(M)}=0$ and a band crossing of FB is enforced by SR at $M$.
In a similar way, one can enumerate all possible SRs for nondegenerate FB in the Lieb lattice.
We find that there are only two cases where the FB is nonsingular: (i) a CLS has $\bb A_0$ SR regardless of which sublattices are occupied and (ii) a CLS occupies all three sublattices and has $\bb D_2$ SR.
Here, SR $\bchi_l$ indicates that the corresponding CLS is centered at Wyckoff position $\bchi$ and has $C_4$ eigenvalue $e^{i\pi l/2}$.
All the other SRs lead to SFBs.

So far, we have discussed the SRs under $C_4$ only.
However, the Lieb lattice has vertical and horizontal mirrors with normal vectors $\hat{x}$ and $\hat{y}$ and two diagonal mirrors with normal vectors $\hat{x} \pm \hat{y}$ as well.
Together with $C_4$, these symmetries form the $C_{4v}$ point group.
Hence, the CLSs in the Lieb lattice can be classified according to their SRs under $C_{4v}$.
In this way, we can first classify all possible SFBs and NSFBs based on their SRs and the information about which sublattices are occupied, and then a FB model with and without band crossing point(s) of FB can be constructed through the KPC scheme.

\subsection{Flat-band models with nonsingular flat bands \label{subsec:NSFB_Lieb}}
As discussed above, the KPC scheme can also be used to construct a NSFB when the condition for the SR-enforced band crossing is avoided.
In this section, we construct FB models in the Lieb lattice which exhibit NSFBs without band crossing points.

\begin{figure}[t!]
\centering
\includegraphics[width=0.42\textwidth]{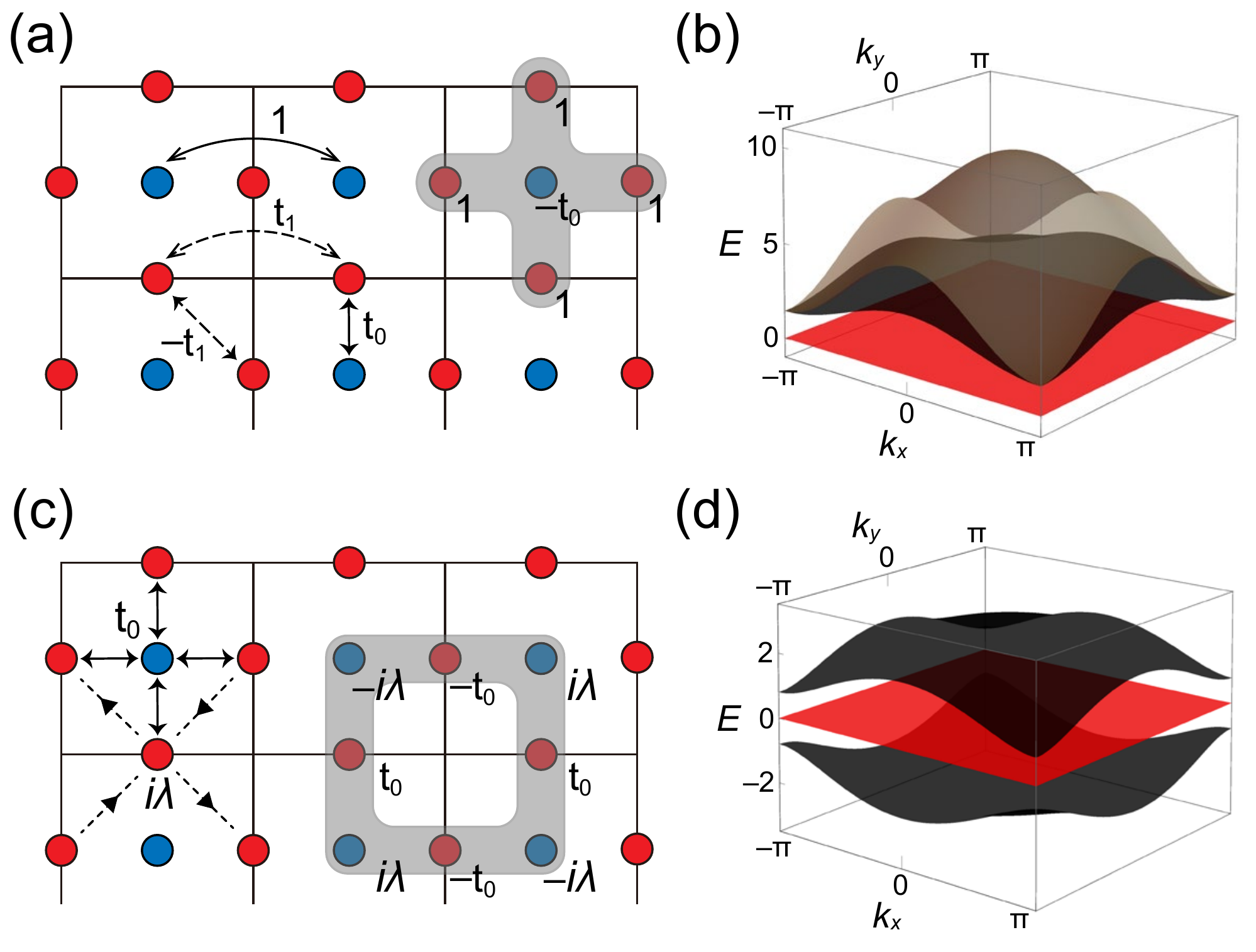}
\caption{
Non-singular flat bands (NSFBs) in the Lieb lattice.
(a) Description of the gapped FB model $H'_{\rm Lieb}(\bk)$.
Arrows denote the hoppings, and the CLS $\ket{w'_{\rm Lieb}(\bR)}$ is drawn in gray region.
The onsite potentials of the blue and red sublattices are equal to $4$ and $t_0^2+2t_1$, respectively.
For clear illustration, we draw a minimal set of hoppings.
Others not shown here can be generated by $C_4$ rotation.
The CLS $\ket{w'_{\rm Lieb}(\bR)}$ has $\bb A_0$ SR.
(b) Band structure of $H'_{\rm Lieb}(\bk)$ for $(t_0,t_1)=(1.2,0.4)$.
(c) Description of the gapped FB model $H''_{\rm Lieb}(\bk)$.
$\lambda$ denotes the strength of spin-orbit coupling.
The CLS $\ket{w'_{\rm Lieb}(\bR)}$ has $\bb D_2$ SR and occupies all three sublattices.
(d) Band structure of $H''_{\rm Lieb}(\bk)$ for $(t_0,\lambda)=(1.0,0.2)$.
}
\label{Fig3}
\end{figure}

First, we consider the case (i) a CLS has $\bb A_0$ SR.
To this end, we choose a specific CLS $\ket{w'_{\rm Lieb}(\bR)}=-t_0 \ket{\bR,1}+\ket{\bR,2}+\ket{\bR-\bb a_1,2}+\ket{\bR,3}+\ket{\bR-\bb a_2,3}$, shown in \fig{Fig3}(a).
The corresponding FT-CLS is given by $\uefL{''}{\rm Lieb}=(-t_0,1+\cm{Q_1},1+\cm{Q_2})$.
Clearly, it is nonsingular as long as $t_0$ is nonzero.
Then, we choose three BMOs $\bskL{'}{1}{\bk}=(0,1+Q_2,-1-Q_1)$, $\bskL{'}{2}{\bk}=(1+Q_2,0,t_0)$, and $\bskL{'}{3}{\bk}=(1+Q_1,t_0,0)$.
For illustration, we construct a simple KPC Hamiltonian,
\ba
H'_{\rm Lieb}(\bk) = t_1 h'_{11}(\bk)+h'_{22}(\bk)+h'_{33}(\bk),
\label{eq:H_gappedLieb}
\ea
where $h'_{ab}= \kbr{\vph'_a(\bk)}{\vph'_b(\bk)}$ ($a,b=1,2,3$).
The band structure for $t_0=1.2$ and $t_1=0.4$ is shown in \fig{Fig3}(b).
As expected from the nonsingularity of $\uefL{''}{\rm Lieb}$, the FB of $H'_{\rm Lieb}(\bk)$ is gapped.
%

Now, we consider the case (ii) where a CLS has $\bb D_2$ SR and occupies the three sublattices.
A relevant model is known as the spin-orbit coupled Lieb model~\cite{weeks2010topological} and its Hamiltonian is given by
\ba
H''_{\rm Lieb}(\bk) = t_0 \bpm
0 & (1+Q_1) & (1+Q_2) \\
(1+\cm{Q_1}) & 0 & g_{\rm soc}(\bk) \\
(1+\cm{Q_2}) & \cm{g_{\rm soc}(\bk)} & 0
\epm,
\label{eq:socLieb_H}
\ea
where $g_{\rm soc}(\bk)=i\frac{\lambda}{t_0}(1-\cm{Q_1})(1-Q_2)$ and  $\lambda$ corresponds to the spin-orbit coupling between the orbitals at the second and third sublattices [\fig{Fig3}(c)].
The band structure for $t_0=1.0$ and $\lambda=0.2$ is shown in \fig{Fig3}(d).
A FB exists without band crossing points.
The corresponding CLS can be obtained from the FT-CLS, $\uefL{''}{\rm Lieb} = (i \lambda (1-Q_1)(1-Q_2),-t_0-t_0 Q_2,t_0+t_0 Q_1)$, and its shape is illustrated in \fig{Fig3}(c).
We note that the spin-orbit coupled Lieb model $H''_{\rm Lieb}(\bk)$ cannot be obtained by the KPC scheme, which indicates that there are a class of FB models indescribable by the KPC.
[Of course, a new FB model having a NSFB with $\bb D_2$ SR can be constructed by choosing the BMOs suitably for $\uefL{''}{\rm Lieb}$ in a similar way as we construct the FB model $H'_{\rm Lieb}(\bk)$ in \eq{eq:H_gappedLieb}.]
Although $H''_{\rm Lieb}(\bk)$ can be expressed as the sum of Kronecker products of BMOs, our assumption that $f_{ab}(\bk)$ in \eq{eq:KPC_Lieb} is a Laurent polynomial is violated.
As detailed in Appendix~\ref{app:exceptional}, this class includes FB models constrained by an antiunitary symmetry $C \circ I_{ST}$, a combination of chiral $C$ and space-time inversion $I_{ST}$ symmetries.

\section{Detailed procedures for Kronecker-product construction \label{sec:KPC}}
In the previous section, the KPC scheme is applied to construct various FB models in the Lieb model.
Here, we establish the detailed procedures for KPC scheme, which are composed of five steps (I-V):
\begin{itemize}
\item \textbf{I.} Choose a CLS $\ket{w(\bR)}$ in a given lattice.
\item \textbf{II.} Check that whether the FT-CLS $\uef{}$ corresponding to $\ket{w(\bR)}$ is singular or nonsingular.
\item \textbf{III.} Set BMOs with respect to $\uef{}$ following the four prescriptions detailed in Section~\ref{subsec:KPC_basis}.
\item \textbf{IV.} Construct a tight-binding Hamiltonian using the BMOs.
\item \textbf{V.} Symmetrize the tight-binding Hamiltonian.
\end{itemize}

\subsection{Conventions \label{subsec:KPC_Conventions}}
Before we explain the detailed procedures of KPC scheme, we clarify our conventions used in this work.
We mainly discuss tight-binding models with nondegenerate FB unless otherwise noted.
(Construction of FB model with degenerate FB is discussed in Sec.~\ref{sec:deg_flat}.)
Such FB models are defined in $d$-dimensional ($d$D) symmorphic lattice with the primitive lattice vectors $\bb a_i$ ($i=1,\dots,d$).
Each unit cell is labeled by lattice vector $\bR$ and consists of $n_{\rm tot}$ sublattice sites $\bR+\bx_\alpha$ ($\alpha=1,\dots,n_{\rm tot}$).
An atomic orbital at the sublattice site $\bR+\bx_\alpha$ is denoted as $\ket{\bR,\alpha}$.
In the tight-binding limit, the atomic orbitals are orthonormal such that  $\brk{\bR,\alpha}{\bR',\beta}=\delta_{\bR,\bR'}\delta_{\alpha\beta}$.

In real space, a tight-binding Hamiltonian is given by
\ba
\hat{H} = \sum_{\bR,\Delta \bR} \sum_{\alpha,\beta} t_{\alpha \leftarrow \beta}(\Delta \bR) \ket{\bR+\Delta \bR, \alpha} \bra{\bR,\beta},
\ea
where $t_{\alpha \leftarrow \beta}(\Delta \bR)$ denotes a hopping parameter between atomic orbitals at $\bR+\Delta \bR+\bx_\alpha$ and $\bR+\bx_\beta$, respectively.
Then, a $n_{\rm tot} \times n_{\rm tot}$ Hamiltonian in momentum space, $H(\bk)_{\alpha\beta}=\bra{\bk,\alpha} \hat{H} \ket{\bk,\beta}$, is obtained by using $\ket{\bk,\alpha} = N_{\rm cell}^{-1/2} \sum_{\bR} e^{i \bk \cdot \bR} \ket{\bR, \alpha}$, where $N_{\rm cell}$ is the number of unit cells. Explicitly,
\ba
H(\bk)_{\alpha\beta} = \sum_{\Delta \bR} t_{\alpha \leftarrow \beta}(\Delta \bR) e^{-i\bk \cdot \Delta \bR}.
\ea
We note that our definition of $\ket{\bk,\alpha}$ is different from the conventional one defined as $\ket{\bk,\alpha}' = N_{\rm cell}^{-1/2} \sum_{\bR} e^{i \bk \cdot (\bR+\bx_\alpha)} \ket{\bR, \alpha}$.
Crucially, both $\ket{\bk,\alpha}$ and $H(\bk)_{\alpha\beta}$ is periodic in the Brillouin zone, i.e. $\ket{\bk+\bb G,\alpha}=\ket{\bk,\alpha}$ and $H(\bk+\bb G)_{\alpha\beta}=H(\bk)_{\alpha\beta}$ for any reciprocal lattice vector $\bb G$.
For this reason, we call $H(\bk)_{\alpha\beta}$ a tight-binding Hamiltonian in the \textit{periodic} basis.
The periodicity in the Brillouin zone (BZ) makes our construction scheme simple.
For example, each element of $H(\bk)$ is simply expressed as a Laurent polynomial in $d$ variables $Q_i \equiv e^{-i\bk \cdot \bb a_i}$ and does not depend on the positions of sublattice sites $\bx_\alpha$.
Also, in the periodic basis, the hopping structure of a given model can be simply read off by using
\ba
t_{\alpha \leftarrow \beta}(\Delta \bR) = \frac{1}{N_{\rm cell}} \, \sum_{\bk} e^{i \bk \cdot \Delta \bR} H(\bk)_{\alpha\beta}
\label{eq:hopping}.
\ea
Some useful formulas related to the tight-binding Hamiltonian in the periodic basis are summarized in Appendix~\ref{app:TB}.

We note one important condition that any CLS $\ket{w(\bR)}$ discussed in this work must be elementary~\cite{hwang2021flat} such that $\ket{w(\bR)}$ cannot be spanned by another CLS $\ket{w'(\bR)}$.
Equivalently, this condition can be rephrased using the corresponding FT-CLS $\uef{}$, which can be obtained from the chosen CLS through a Fourier transform [see \eqs{eq:CLS_def1} and \eqref{eq:CLS_def2}]: There must be no common divisor polynomial of all elements in $\uef{}$ except \textit{monomial} of the form $c \, \prod_i Q_i^{n_i}$ with $c \in \mathbb{C}$ and $n_i \in \mathbb{Z}$.
We call this condition an \textit{irreducibility}.
For example, $\ket{\hat{u}(\bk)}=(1+Q_1+Q_2^{-1}+Q_1 Q_2^{-1},1+Q_1$) is not a proper FT-CLS, since there is a common divisor polynomial $(1+Q_1)$.

\subsection{Step I: Choose a compact localized state \label{subsec:KPC_CLS}}
We begin our construction scheme by choosing a CLS $\ket{w(\bR)}$ which takes the form of \eq{eq:CLS_def1}.
When we are interested in nondegenerate FB, a CLS must be centered at the maximal Wyckoff position $\bchi$ with unit multiplicity 1.
The site symmetry $G_\bchi$ of $\bchi$ leaves $\bchi$ invariant up to lattice translation.
Then, we choose a CLS with a definite SR for each symmetry element $\sg=\{O_\sg|\bb \delta_\sg\} \in G_\bchi$.
Here, $\{O_\sg|\bb \delta_\sg\}$ denotes the action of $\sg$ on real-space coordinates $\bb r$ as $\sg: \bb r \rightarrow O_\sg \bb r + \bb \delta_\sg$ where $O_\sg$ is a $d \times d$ orthogonal matrix.
The SRs of CLS are determined by $\bchi$ and symmetry eigenvalues $\xi_\sg$ for $\sg$.
According to SRs, the CLS transforms as
\ba
\hat{\sg} \ket{w(\bR)} = \ket{w(\bR_{(\sg,\bchi)})} \xi_\sg,
\label{eq:SR_CLS}
\ea
where $\bR_{(\sg,\bchi)} = O_\sg \bR + O_\sg \bchi - \bchi + \bb \delta_\sg$.
Note that, in the case of $\ket{w_{\rm Lieb}(\bR)}$, $\sg=C_4$, $\bchi=\bb D$, and $\xi_\sg=-1$.
For a detailed discussion on SR of CLS, see \Rf{hwang2021flat}.

\subsection{Step II: Find a singularity of FT-CLS \label{subsec:KPC_singular}}
When a FT-CLS $\uef{}$ is singular at $\bk_*$, i.e. $\ket{\hat{u}(\bk_*)}=0$, a FB must have a band crossing with other bands at the singular point $\bk_*$~\cite{rhim2019classification,hwang2021flat}.
Hence, one can determine whether a chosen CLS leads to a band crossing between the flat and other bands or not, by simply identifying the singular points of $\uef{}$ even before the construction of the Hamiltonian.

As discussed in Sec.~\ref{subsec:singular_Lieb}, a class of SR enforces band crossing point(s) of FBs.
Here, we briefly review the condition for having such band crossing points enforced by SR, which is rigorously described in \Rf{hwang2021flat}.
The SR of CLS determines the symmetry transformation of FT-CLS.
It can be shown that \eq{eq:SR_CLS} leads to
\ba
U_{\sg}(\bk) \uef{} = \ket{\hat{u}(O_\sg \bk)} \, \xi_\sg(\bk)
\label{eq:SR_FT-CLS}
\ea
where $\xi_\sg(\bk) = \xi_\sg e^{-i O_\sg \bk \cdot (O_\sg \bchi - \bchi + \bb \delta_\sg)}$ and $U_\sg(\bk)$ denotes a symmetry operator for $\sg$.
Note that the detailed form of $U_\sg(\bk)$ depends on how $\sg$ acts on atomic orbitals in a given lattice (see Appendix~\ref{app:TB}).
Crucially, when \eq{eq:SR_FT-CLS} imposes $\ket{\hat{u}(\bk_\sg)}=0$ at a high-symmetry point satisfying $\bk_\sg = O_\sg \bk_\sg$ (mod $\bb G$), a FB is enforced to be a SFB.
Note also that in order to determine whether $\uef{}$ is singular or not, it must also be considered that which sublattices are occupied by the CLS.
Recall that a CLS with $\bb D_2$ SR in the Lieb lattice corresponds to SFB when the first sublattice is not occupied, however, it corresponds to NSFB when all three sublattices are occupied, as discussed in Sec.~\ref{subsec:singular_Lieb}.

\subsection{Step III: Set basis molecular orbitals \label{subsec:KPC_basis}}
Next, we set BMOs $\bsk{a}$ ($a=1,\dots,n_B$).
The BMOs are orthogonal to the given FT-CLS $\uef{}$ such that $\brk{\vph_a(\bk)}{\hat{u}(\bk)}=0$.
We require that each element of BMOs is a Laurent polynomial to ensure that the resulting KPC Hamiltonian involves hoppings with a finite range.
Also, each BMO must satisfy the irreducibility defined in Sec.~\ref{subsec:KPC_Conventions}.

In general, the BMOs do not have to be orthogonal to each other and are not energy eigenstates.
For this reason, the number of BMOs $n_B$ can be larger than $n_{\rm tot}-1$.
Although any choice of BMOs gives a FB model, we list the following four prescriptions for choosing a set of BMOs to have a less complicated hopping structure or to reconstruct known FB models through the KPC scheme.

Briefly, we comment on the role of each prescription.
The first prescription is about the number of BMOs.
In the case of a generic FT-CLS, BMOs can be found according to the second prescription.
The last two prescriptions determine how to find BMOs when the elements of FT-CLS have specific relationships between them.
The BMOs determined by the last two prescriptions have smaller shapes than those from the second prescription.
The smaller the shape of BMOs, the smaller the range of hoppings [see also \eq{eq:KPC_hopping}].

\subsubsection{Prescription 1 \label{subsubsec:prescription1}}
First, in order to have a nondegenerate FB, we must set at least $(n_{\rm tot}-1)$ number of BMOs.
In fact, the number of BMOs ($n_B$) and the number of independent (complex-valued) BMOs at $\bk$ [$N_B(\bk)$] must be distinguished.
That is, $N_B(\bk)$ is equal to or less than $n_B$, and it can vary depending on $\bk$.
For example, consider three BMOs, $\bsk{1}=(0,-1-Q_3,1+Q_2)$, $\bsk{2}=(1+Q_3,0,-1-Q_1)$, and $\bsk{3}=(-1-Q_2,1+Q_1,0)$.
In this case, the number of BMOs is $n_B=3$.
Since $(1+Q_1)\bsk{1}+(1+Q_2)\bsk{2}+(1+Q_3)\bsk{3}=0$ and $\bsk{1,2,3} \ne 0$ at $(Q_1,Q_2,Q_3)\ne(-1,-1,-1)$, the number of independent complex-valued vectors $\{\bsk{a}\}$ is $N_B(\bk)=2$.
However, at $(Q_1,Q_2,Q_3)=(-1,-1,-1)$, all the BMOs vanish and hence $N_B(\bk)=0$.
Hence, a FB model constructed from these BMOs exhibit a single FB with a band crossing at $(Q_1,Q_2,Q_3)=(-1,-1,-1)$ with three-fold degeneracy.
This observation can be generalized to arbitrary cases: A KPC Hamiltonian exhibits at least $[n_{\rm tot}-{\rm max} \, N_B(\bk)$] number of FBs at zero energy, and the FBs have a band crossing with dispersive bands at the singular point $\bk_*$ with at least [$n_{\rm tot}-N_B(\bk_*)]$-fold degeneracy.
Hence, we can also construct a FB model with degenerate FB by suitably choosing the detailed expression and the number of BMOs, as discussed in Sec.~\ref{sec:deg_flat}.

\subsubsection{Prescription 2 \label{subsubsec:prescription2}}
Second, we define \textit{canonical} BMOs.
Consider a FT-CLS whose elements have no particular relationship among them such as $\uef{}_1=\uef{}_2$ or $\uef{}_1=0$.
For such FT-CLS, the canonical BMOs can be set as follows: First, choose two different sublattice indices $a$ and $b$, i.e. $a\ne b$.
Then, a canonical BMO $\ket{\phi_{(a,b)}(\bk)}$ has its elements as $\ket{\phi_{(a,b)}(\bk)}_{a}=-\cm{\uef{}_b}$, $\ket{\phi_{(a,b)}(\bk)}_{b}=\cm{\uef{}_a}$, and $\ket{\phi_{(a,b)}(\bk)}_{\alpha \ne a,b}=0$.
Since each canonical BMO is made by choosing two different elements in a given FT-CLS, the number of canonical BMOs is equal to $n_B={}_{n_{\rm tot}}\mathrm{C}_{2}$ where ${}_{n}\mathrm{C}_{m}$ denotes the number of combinations of $m$ from $n$.
For example, when $n_{\rm tot}=3$, we set three canonical BMOs: $\ket{\phi_{(1,2)}(\bk)}=(-\cm{u_2},\cm{u_1},0)$, $\ket{\phi_{(1,3)}(\bk)}=(-\cm{u_3},0,\cm{u_1})$, and $\ket{\phi_{(2,3)}(\bk)}=(0,-\cm{u}_3,\cm{u_2})$, where $\uef{}_a=u_a$.

\subsubsection{Prescription 3 \label{subsubsec:prescription3}}
Third, we identify which sublattices the CLS does not occupy and label them as $\alpha_\varnothing$.
Accordingly, the corresponding elements in $\uef{}$ are identically zero, i.e. $\uef{}_{\alpha_\varnothing}=0$.
Suppose that the number of unoccupied sublattices is equal to $n_\varnothing$, and a set of unoccupied sublattices $\{\alpha_\varnothing\}$ can be expressed as an ordered set $\left\{(\alpha_\varnothing)_1,\dots,(\alpha_\varnothing)_{n_\varnothing}\right\}$.
In this case, our set of BMOs must include $n_\varnothing$ BMOs $\bsk{I=1,\dots,n_\varnothing}$ such that $\bsk{I}_\alpha=\delta_{(\alpha_\varnothing)_I,\alpha}$.
As an instance, let us consider $\uef{\rm Lieb}=(0,-1-Q_2,1+Q_1)$ in the Lieb lattice where the first sublattice is unoccupied.
As only one sublattice is unoccupied, $\{\alpha_\varnothing\}=\{1\}$ and $n_\varnothing=1$.
Then, using the prescription, we can set $\bsk{1}=(1,0,0)$.
As a second BMO, we can choose $\bsk{2}=(0,1+Q_1^{-1},1+Q_2^{-1})$ which is nothing but a canonical BMO $\ket{\phi_{(2,3)}(\bk)}$ [see \eq{eq:Lieb_basis}].
This prescription is useful when a given lattice is bipartite and two mutually disjoint sets of sublattices have unequal cardinalities.
Note that three sublattices of Lieb lattice is divided into two disjoint sets, blue and red sublattices with cardinalities one and two respectively, as shown in \fig{Fig2}(a).

\subsubsection{Prescription 4 \label{subsubsec:prescription4}}
Fourth, suppose that some elements of $\uef{}$ have mutual relationships.
Such relationships typically take the following form,
\ba
\sum_{\alpha=1}^{n_{\rm tot}} \, M_\alpha(\bk) \uef{}_\alpha = 0,
\label{eq:u_relation}
\ea
where $M_\alpha(\bk)$ is a monomial in $Q_{1,\dots,d}$, $M_\alpha(\bk)=(2+i) \times Q_1\cm{Q_2}$ for example.
This situation naturally occurs in FB models on regular lattices such as split and line graphs.
For the relationship in \eq{eq:u_relation}, we choose a relevant BMO as
\ba
\bsk{\rm R}=\left(\cm{M_1(\bk)},\dots,\cm{M_{n_{\rm tot}}(\bk)}\right).
\ea
Then, \eq{eq:u_relation} is identical to the orthogonality, $\brk{\vph_{\rm R}(\bk)}{\hat{u}(\bk)}=0$, which is a necessary condition for $\bsk{\rm R}$ being a BMO.
In fact, $\uef{}_{\alpha_\varnothing}=0$, which is relevant to the third prescription, is the simplest case of \eq{eq:u_relation}.
Especially, we follow the fourth prescription to construct a FB model in the kagome lattice in Sec.~\ref{subsec:ex_kagome}.

For clear illustration, let us consider $\uef{}=(u_1,u_2,u_3,u_4)$ with $u_{\alpha}=\uef{}_\alpha$ ($\alpha=1,2,3,4$).
When there is no relationship between $u_{1,2,3,4}$, we can find six canonical BMOs following the second prescription: $\ket{\phi_{(1,2)}(\bk)}=(-\cm{u_2},\cm{u_1},0,0)$, $\ket{\phi_{(1,3)}(\bk)}=(-\cm{u_3},0,\cm{u_1},0)$, $\ket{\phi_{(1,4)}(\bk)}=(-\cm{u_4},0,0,\cm{u_1})$, $\ket{\phi_{(2,3)}(\bk)}=(0,-\cm{u_3},\cm{u_2},0)$, $\ket{\phi_{(2,4)}(\bk)}=(0,-\cm{u_4},0,\cm{u_2})$, and $\ket{\phi_{(3,4)}(\bk)}=(0,0,-\cm{u_4},\cm{u_3})$.
On the other hand, if there are some relationships between $u_{1,2,4}$ such as $u_4=0$ and $u_1+u_2=0$ for example, we can apply the third and fourth prescriptions for the former and the latter, respectively.
Since $u_4=0$, the third prescription determines a BMO $\bsk{1}=(0,0,0,1)$.
Then, taking account of $u_1+u_2=0$, we set $\bsk{2}=(1,1,0,0)$ following the fourth prescription.
When there are $n_R$ number of such relationships between the elements of $\uef{}$, the number of independent elements in $\uef{}$ reduces to $n_{\rm tot}-n_R$.
Thus, it is sufficient to additionally include ${}_{n_{\rm tot}-n_R}\mathrm{C}_2$ canonical BMOs.
In the example above, $(n_{\rm tot},n_R)=(4,2)$ and ${}_{n_{\rm tot}-n_R}\mathrm{C}_2={}_{4-2}\mathrm{C}_2=1$.
Hence, we need to choose one canonical BMO $\bsk{3}=\ket{\phi_{(1,3)}(\bk)}$, in addition to $\bsk{1}$ and $\bsk{2}$.
Note that six canonical BMOs $\ket{\phi_{(a,b)}(\bk)}$ ($a,b=1,\dots,4$ and $a\ne b$) are expressed as a linear combination of three BMOs $\bsk{1,2,3}$ with Laurent polynomial coefficients:
$\ket{\phi_{(1,2)}(\bk)}=\cm{u_1} \bsk{2}$, $\ket{\phi_{(1,3)}(\bk)}=\bsk{3}$, $\ket{\phi_{(1,4)}(\bk)}=\cm{u_1} \bsk{1}$, $\ket{\phi_{(2,3)}(\bk)}=-\bsk{3}-\cm{u_3}\bsk{2}$, $\ket{\phi_{(2,4)}(\bk)}=\cm{u_2} \bsk{1}$, and $\ket{\phi_{(3,4)}(\bk)}=\cm{u_3} \bsk{1}$.
Hence we take the set of BMOs $\{\bsk{1,2,3}\}$ instead of $\{\ket{\phi_{(a,b)}(\bk)}\}$, as the latter is spanned by the former.

Based on the above observation, we note that the fourth prescription can be summarized in a more general but abstract form.
Suppose we can find two different sets of BMOs $\{\bsk{a=1,\dots,n_B}\}$ and $\{\bskL{'}{a'=1,\dots,n'_B}{\bk}\}$ and $n'_B>n_B$.
If $\{\bskL{'}{a'}{\bk}\}$ is represented as a linear combination of $\{\bsk{a}\}$ with Laurent polynomial coefficients, i.e. $\bskL{'}{a'}{\bk} = \sum_{a=1}^{n_B} \, L_{a'a}(\bk) \bsk{a}$ where $L_{a'a}(\bk)$ is a Laurent polynomial in $Q_i$, we choose $\{\bsk{a}\}$ as our BMOs.

\subsection{Step IV: Construct a KPC Hamiltonian \label{appsec:KPC_Htb}}
Now, with the BMOs $\bsk{a}$ ($a=1,\dots,n_B$), we construct a tight-binding Hamiltonian in periodic basis as
\ba
H_{\rm KPC}(\bk) = \sum_{a,b=1}^{n_B} \, f_{ab}(\bk) \, h_{ab}(\bk),
\label{eq:KPC}
\ea
where $h_{ab}(\bk)=h_{ba}(\bk)^{\dg}=\bsk{a}\bsb{b}$ and the Laurent polynomials $f_{ab}(\bk)$ in $Q_i$.
Note that $f_{ab}(\bk)=\cm{f_{ba}(\bk)}$ to ensure hermiticity of $H_{\rm KPC}(\bk)$.
In this perspective, the existence of flat band at zero energy is obviously due to the orthogonality, $\brk{\vph_a(\bk)}{\hat{u}(\bk)}=0$.
We note that each element of $H_{\rm Lieb}(\bk)$ is a Laurent polynomial, thus $H_{\rm Lieb}(\bk)$ contains only finite-ranged hoppings.

The hopping structure can be readily inferred from the shape of BMOs and the form of $f_{ab}(\bk)$ with the help of \eq{eq:hopping}.
Explicitly, a hopping parameter is expressed as
\ba
t_{\alpha \leftarrow \beta}(\Delta \bR) =& \sum_{a,b}\sum_{\bR',\bR_{(ab)}} F[\bR_{(ab)}] S^{(a)}_\alpha (\widetilde{\bR}) \cm{S^{(b)}_\beta(\bR')},
\label{eq:KPC_hopping}
\ea
where $\widetilde{\bR}=\bR'-\bR_{(ab)}+\Delta \bR$, and $F[\bR_{(ab)}]$ and $S^{(a)}_\alpha(\bR)$ are given by
\bg
F[\bR_{(ab)}] = \frac{1}{N_{\rm cell}} \sum_\bk \, f_{ab}(\bk) \, e^{i \bk \cdot \bR_{(ab)}}, \\
S^{(a)}_\alpha(\bR) = \frac{1}{N_{\rm cell}} \sum_\bk \, \bsk{a}_\alpha \, e^{i \bk \cdot \bR},
\eg
respectively.
Equation~\eqref{eq:KPC_hopping} implies that a range of hopping $\Delta \bR$ between $\alpha$th and $\beta$th orbitals are determined by the shapes of BMOs.
This can be clearly seen if $f_{ab}(\bk)=1$ is assumed for simplicity, where \eq{eq:KPC_hopping} is reduced to $t_{\alpha \leftarrow \beta}(\Delta \bR) = \sum_{a,b}\sum_{\bR'} S^{(a)}_\alpha (\bR'+\Delta \bR) \cm{S^{(b)}_\beta(\bR')}$.
The hopping parameter $t_{\alpha \leftarrow \beta}(\Delta \bR)$ becomes zero when $\bR'+\Delta \bR$ is outside the shape of $\bskL{}{a}{\bk}$, i.e. $S^{(a)}_\alpha (\bR'+\Delta \bR)=0$, or $\bR'$ is outside the shape of $\bskL{}{b}{\bk}$.

\subsection{Step V: Symmetrization \label{subsec:KPC_symmetrization}}
The chosen CLS has SRs under symmetry group $G_\bchi$ as discussed in Sec.~\ref{subsec:KPC_CLS}.
However, if our choice of BMOs or $f_{ab}(\bk)$ does not respect $G_\bchi$, the resulting KPC Hamiltonian is not symmetric under $G_\bchi$.
In such a case, we symmetrize $H_{\rm KPC}(\bk)$ through the symmetrization algorithm~\cite{gresch2018automated}.
In Appendix~\ref{app:symmetrization}, we review the symmetrization algorithm for tight-binding Hamiltonian~\cite{gresch2018automated}, and we prove that the FT-CLSs remain unchanged after the symmetrization.
When the symmetrization is applied, additional BMOs required to respect $G$ are generated, or $f_{ab}(\bk)$ are modified such that $H_{\rm KPC}(\bk)$ is symmetric under $G_\bchi$.

\section{More examples of flat-band models \label{app:examples}}
In this section, we construct FB models in various lattices based on the KPC scheme.
We note that the hopping structures of FB models discussed in the following can be read off by using \eq{eq:hopping}.

\subsection{NSFB in 1D inversion-symmetric lattice \label{subsec:ex_1D}}
\begin{figure}[t!]
\centering
\includegraphics[width=0.45\textwidth]{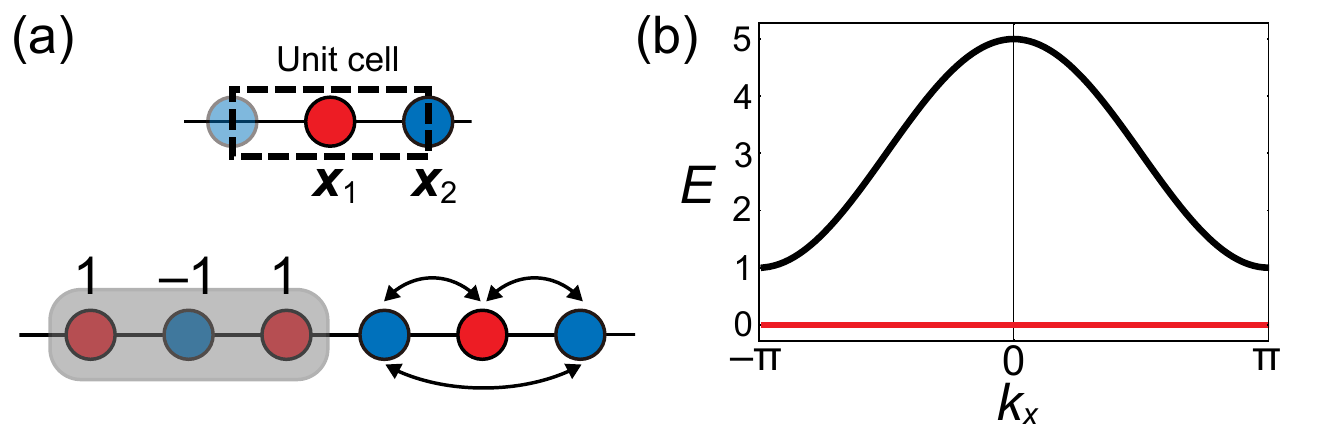}
\caption{
Flat-band model in 1D inversion-symmetric lattice.
(a) Two sublattices are located at $x_1=0$ and $x_2=1/2$.
Arrows denote the hoppings with the strength 1.
Red and blue sublattices have different on-site energies, 1 and 2, respectively.
The CLS is illustrated in the gray region.
The numbers near each sublattice denote the amplitude of the CLS.
(b) Band structure of $H_{\rm 1D}(k_x)$.
A gapped FB (red) appear at zero energy.
}
\label{Fig4}
\end{figure}

Consider an inversion-symmetric lattice system in 1D where two sublattices are located at $x_1=0$ and $x_2=1/2$, respectively [\fig{Fig4}(a)].
Note that $x_1$ and $x_2$ correspond to the maximal Wyckoff positions.
An $s$ orbital is located at each sublattice.
Then, the symmetry operator for inversion $I$ is given by $U_{I}(k_x)={\rm Diag} (1,Q_1)$, since $\hat{I} \ket{R_x,1} = \ket{-R_x,1}$ and $\hat{I} \ket{R_x,2} = \ket{-R_x-1,2}$ where $R_x$ denotes the unit cell index (see Appendix~\ref{app:TB}).
%

First, we consider a CLS that transforms as an $s$ orbital centered at $x_2$.
For such a FB, the CLS and FT-CLS must satisfy
\bg
\hat{I} \, \ket{w(R_x)} = \ket{w(-R_x-1)}, \\
U_I(k_x) \, \ket{\hat{u}(k_x)} = \ket{\hat{u}(-k_x)} \, Q_1,
\eg
where $Q_1=e^{-ik_x}$.
Among many possible choices, we choose
\bg
\ket{w(R_x)} = \ket{R_x,1} + \ket{R_x+1,1} - \ket{R_x,2}, \\
\ket{\hat{u}(k_x)} = (1+Q_1,-1),
\eg
and set a canonical BMO $\ket{\vph(k_x)} = (1,1+\cm{Q_1})$.
Since $\ket{\hat{u}(k_x)}$ is nonzero everywhere and thus nonsingular, we expect that the resulting FB model has NSFB.
Following the KPC scheme, we obtain a FB model described by
\ba
H_{\rm 1D}(k_x) &= \ket{\vph(k_x)} \bra{\vph(k_x)} \nn \\
&= \bpm 1 & 1+Q_1 \\ 1+\cm{Q_1} & 2+Q_1+\cm{Q_1} \epm.
\ea
The detailed description of this model is shown in \fig{Fig4}(a).
The band structure shown in \fig{Fig4}(b) exhibits an isolated and gapped FB, as expected from the fact that there is no SFB in 1D~\cite{rhim2019classification}.

\subsection{SFB in the kagome lattice \label{subsec:ex_kagome}}
\begin{figure}[t!]
\centering
\includegraphics[width=0.45\textwidth]{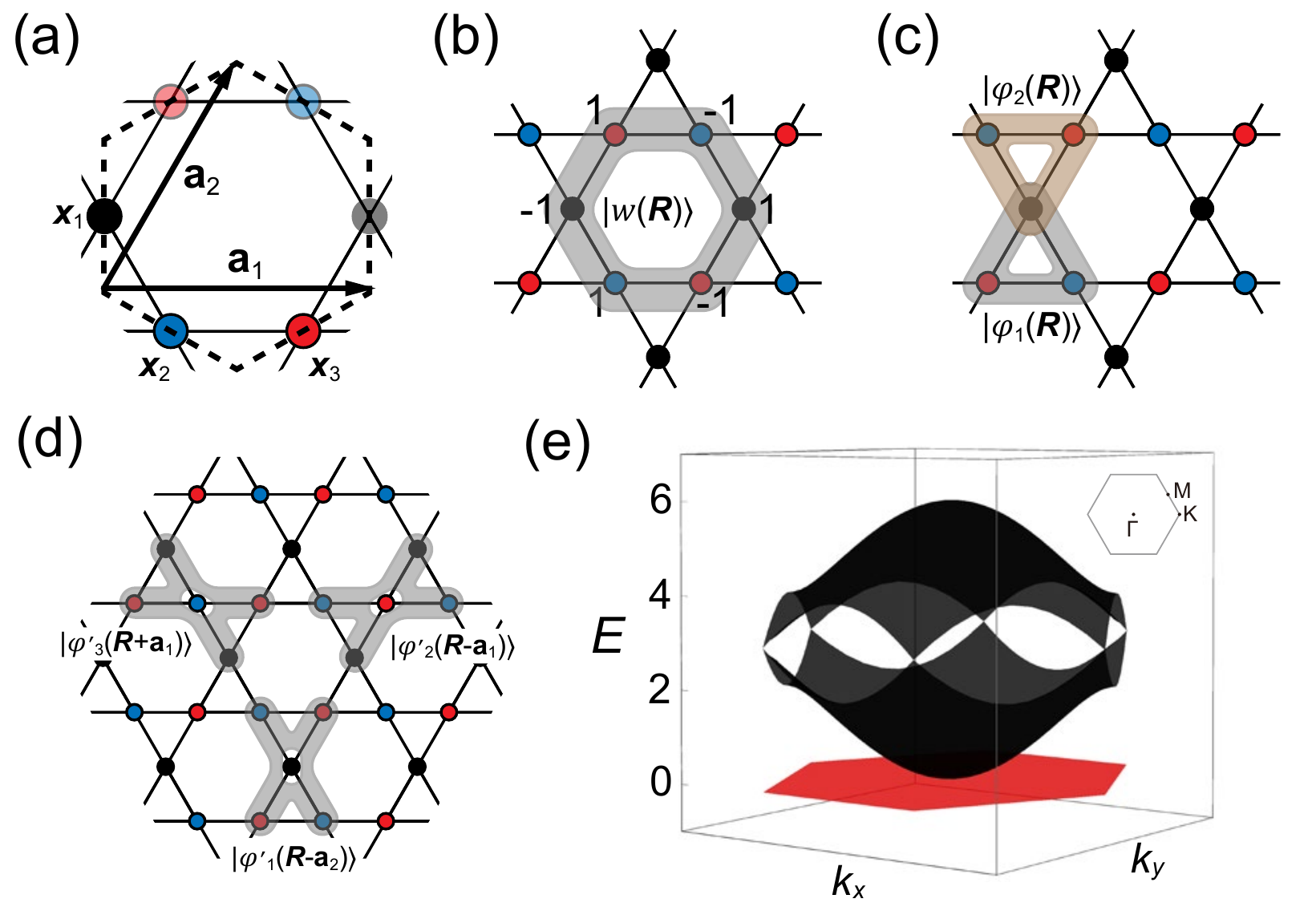}
\caption{
Flat-band model in the kagome lattice.
(a) Three sublattices are represented by black, blue, and red circles.
(b) The CLS is illustrated in the gray region.
The numbers near each sublattice denote the amplitude of the CLS.
(c),(d) Schematic illustration of (c) $\bskL{}{1,2}{\bR}$ and (d) $\bskL{'}{1,2,3}{\bR}$.
(e) The band structure of $H_{\rm kagome}(\bk)$.
A FB has a band crossing at $\Gamma$.
}
\label{Fig5}
\end{figure}

The kagome lattice is composed of three sublattices.
The sublattices are located at $\bx_1=-\frac{1}{2}\bb a_1$, $\bx_2=-\frac{1}{2}\bb a_2$ and $\bx_3=\frac{1}{2}\bb a_1-\frac{1}{2}\bb a_2$ with the primitive lattice vectors $\bb a_1=(1,0)$ and $\bb a_2=(1/2,\sqrt{3}/2)$ [\fig{Fig5}(a)].
Here, we focus on $C_6$ symmetry whose symmetry operator is given by
\ba
U_{C_6}(\bk) = \bpm 0 & 0 & Q_1 \cm{Q_2} \\ 1 & 0 & 0 \\ 0 & 1 & 0 \epm,
\ea
where $Q_i=e^{-i k_i}$ and $k_i=\bk \cdot \bb a_i$ with $i=1,2$.
%

Now, we consider a CLS shown in \fig{Fig5}(b), which corresponds to
\ba
\ket{w(\bR)} =& -\ket{\bR, 1} + \ket{\bR, 2} - \ket{\bR, 3} + \ket{\bR+\bb a_1, 1} \nn \\
&- \ket{\bR+\bb a_2, 2} + \ket{\bR-\bb a_1 + \bb a_2, 3},  \\
\uef{} =& (-1+Q_1,1-Q_2,-1+\cm{Q_1}Q_2).
\ea
The CLS is centered at the $C_6$ rotation center and has  $C_6$ symmetry eigenvalue $\xi_{C_6}=-1$.
Hence, the CLS and FT-CLS transform as
\bg
\hat{C_6} \, \ket{w(\bR)} = - \ket{w(O_{C_6}\bR)}, \nn \\
U_{C_6}(\bk) \uef{} = - \ket{\hat{u}(O_{C_6} \bk)},
\eg
where $(k_1,k_2)$ and $(Q_1,Q_2)$ transform under $C_6$ as $(k_1,k_2) \rightarrow (k_1-k_2,k_1)$ and $(Q_1,Q_2) \rightarrow (Q_1\cm{Q_2},Q_1)$.
This SR of the CLS always leads to a singular FT-CLS at the BZ center $\Gamma=(0,0)$, and hence the corresponding FB has SR-enforced band crossing.
Indeed, $\uef{}=(0,0,0)$ at $\Gamma$.

Now, we set the BMOs with respect to $\uef{}$.
As explained in Sec.~\ref{subsec:KPC_basis}, three canonical BMOs $\{\bskL{'}{a'=1,2,3}{\bk}\}$ can be chosen:
\ba
\bskL{'}{1}{\bk}&=(0,-1+Q_1 \cm{Q_2},\cm{Q_2}-1), \nn \\
\bskL{'}{2}{\bk}&=(1-Q_1\cm{Q_2},0,-1+\cm{Q_1}), \nn \\ \bskL{'}{3}{\bk}&=(1-\cm{Q_2},1-\cm{Q_1},0).
\ea
In fact, $\bskL{'}{2,3}{\bk}$ are generated by acting $C_6$ on $\bskL{'}{1}{\bk}$: $\bskL{'}{2}{O_{C_6}\bk}=U_{C_6}(\bk)\bskL{'}{1}{\bk}$ and $\bskL{'}{3}{O_{C_6}\bk}=U_{C_6}(\bk)\bskL{'}{2}{\bk}$.
Their real-space representations $\bskL{'}{1,2,3}{\bR}$ are shown in \fig{Fig5}(d).
Although FB models can be constructed by using the BMOs $\{\bskL{'}{1,2,3}{\bk}\}$, a relationship between the elements of FT-CLS,
\ba
\uef{}_1+\uef{}_2+Q_1\uef{}_3=0,
\label{eq:relation_kagome}
\ea
indicates that we can find another set of BMOs,
\ba
\bsk{1}=(1,1,\cm{Q_1}), \quad \bsk{2}=(1,\cm{Q_1}Q_2,\cm{Q_1}Q_2).
\ea
Note that \eq{eq:relation_kagome} can be expressed as $\brk{\vph_1(\bk)}{\hat{u}(\bk)}=0$.
The shapes of $\bsk{1,2}$ are illustrated in \fig{Fig5}(c).
Comparing \figs{Fig5}(c) and \hyperref[Fig5]{5}(d), we note that the shapes of $\bsk{1,2}$ are spread over a smaller number of sublattice sites than those of $\bskL{'}{1,2,3}{\bk}$.
Crucially, $\{\bskL{'}{1,2,3}{\bR}\}$ can be represented as a linear combination of $\bsk{1}$ and $\bsk{2}$ with Laurent polynomial coefficients:
\ba
\bskL{'}{1}{\bk} &= -\cm{Q_1}Q_2 \, \bsk{1} + \cm{Q_1}Q_2 \bsk{2}, \nn \\
\bskL{'}{2}{\bk} &= \bsk{1} - Q_1 \, \cm{Q_2}\bsk{2}, \nn \\
\bskL{'}{3}{\bk} &= \bsk{1} - \cm{Q_2} \, \bsk{2}.
\ea
Hence, we choose $\bsk{1,2}$ as the proper BMOs following the fourth prescription in Sec.~\ref{subsec:KPC_basis}.
The BMOs $\bsk{1,2}$ coincide with the molecular orbitals introduced in Refs.~\cite{bilitewski2018disordered,mizoguchi2019molecular,mizoguchi2020systematic}.

The simplest KPC Hamiltonian is constructed by choosing $f_{11}(\bk)=f_{22}(\bk)=1$ and $f_{12}(\bk)=0$:
\ba
H_{\rm kagome}(\bk) &= \sum_{a,b=1}^2 \, f_{ab}(\bk) \bsk{a} \bsb{b} \nn \\
&= \bpm 2 & 1 + Q_1\cm{Q_2} & Q_1+Q_1\cm{Q_2} \\ c.c. & 2 & 1+Q_1 \\ c.c. & c.c. & 2 \epm,
\ea
which corresponds to the well-known FB model in the kagome lattice with the nearest neighbor hoppings.
We note that $f_{11}(\bk)=f_{22}(O_{C_6}\bk)$ and $f_{22}(\bk)=f_{11}(O_{C_6}\bk)$ should be satisfied to preserve $C_6$ since $U_{C_6}(\bk) \bsk{1} = \cm{Q_2} \, \bskL{}{2}{O_{C_6}\bk}$ and $U_{C_6}(\bk) \bsk{2} = \bskL{}{1}{O_{C_6}\bk}$.
The band structure is shown in \fig{Fig5}(e).
A FB has a band crossing at $\Gamma$, as expected from the singularity of FT-CLS.
One can construct more complicated FB models through the KPC scheme by using more general $f_{ab}(\bk)$.

\subsection{SFB in 3D cubic lattice}
\begin{figure}[t!]
\centering
\includegraphics[width=0.42\textwidth]{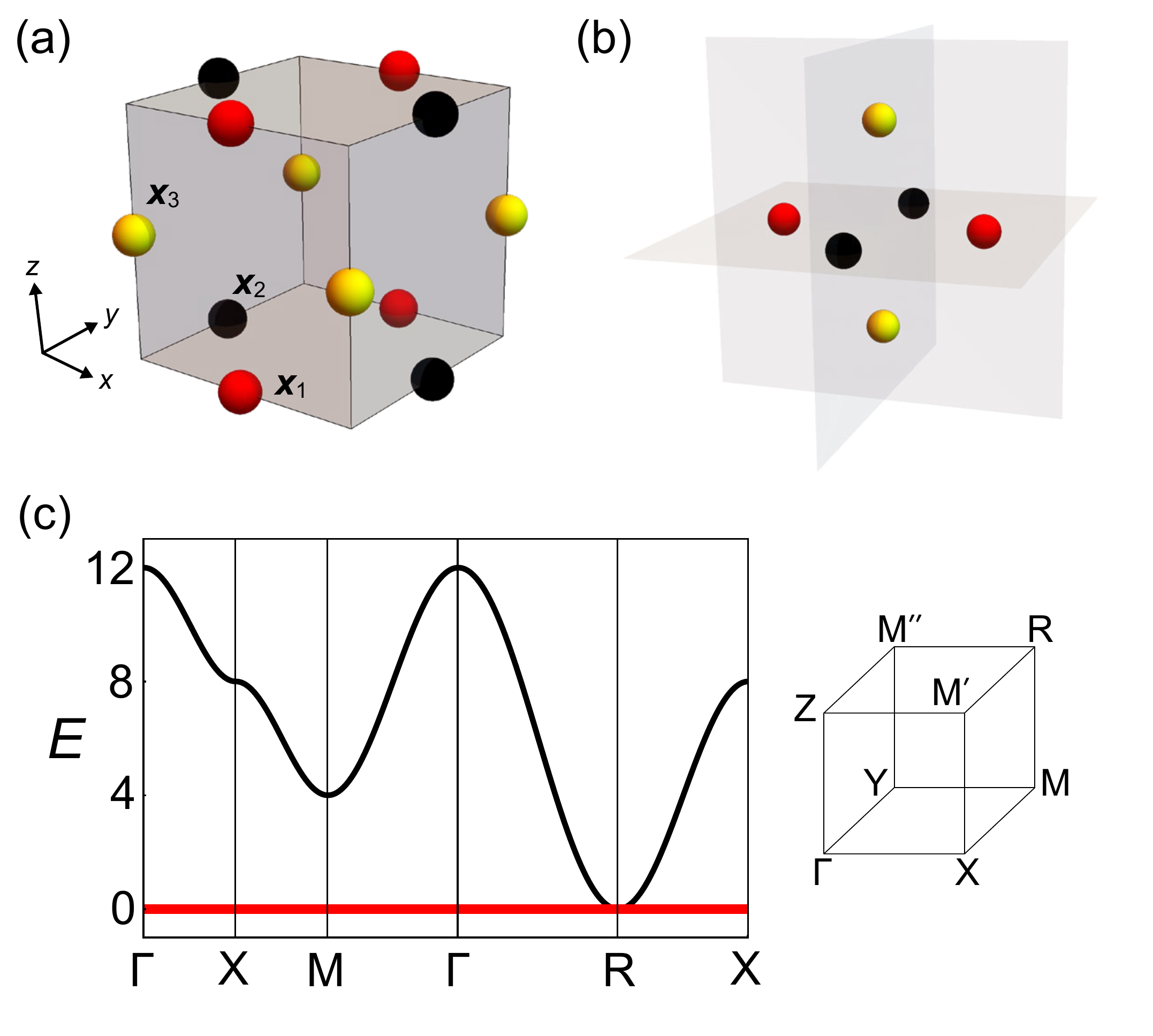}
\caption{
Flat-band model in 3D cubic lattice.
(a) Sublattices are located at $\bx_1=(1/2,0,0)$, $\bx_2=(0,1/2,0)$, and $\bx_3=(0,0,1/2)$, respectively.
(b) Schematic illustration of the CLS.
The amplitudes of the CLS at all sublattices are 1. 
(c) The band structure of $H_{\rm 3D}(\bk)$.
The dispersive bands (black) are two-fold degenerate.
A FB (red) has band crossing at $R$.
The three-fold degeneracy at the band crossing point corresponds to three-dimensional irreducible representation of cubic symmetry group.
}
\label{Fig6}
\end{figure}

We present a FB model in 3D cubic-symmetric lattice.
A unit cell of this lattice is composed of three sublattices located at $\bx_1=(1/2,0,0)$, $\bx_2=(0,1/2,0)$, and $\bx_3=(0,0,1/2)$ [\fig{Fig6}(a)].
The primitive lattice vectors are simply given by $\bb a_1=(1,0,0)$, $\bb a_2=(0,1,0)$, and $\bb a_3=(0,0,1)$.
%

Among symmetry elements in cubic symmetry, let us focus on $C_{2x}$, $C_{2y}$, and $C_{3[111]}$:
\bg
O_{C_{2x}}={\rm Diag}(1,-1,-1), \quad U_{C_{2x}}(\bk) = {\rm Diag}(1,Q_2,Q_3), \nn \\
O_{C_{2y}}={\rm Diag}(-1,1,-1), \quad U_{C_{2y}}(\bk) = {\rm Diag}(Q_1,1,Q_3), \nn \\
O_{C_{3[111]}}=U_{C_{3[111]}}(\bk)=\bpm 0 & 0 & 1 \\ 1 & 0 & 0 \\ 0 & 1 & 0 \epm.
\eg
Taking into account these symmetries, let us consider a CLS $\ket{w(\bR)}$ and the corresponding FT-CLS $\uef{}$ with the following SR:
\bg
\hat{C}_{2x,y} \ket{w(\bR)} = \ket{w(O_{2x,y} \bR)}, \nn \\
U_{C_{2x,y}}(\bk) \uef{} = \ket{\hat{u} (O_{C_{2x,y}}\bk)}, \nn \\
\hat{C}_{3[111]} \ket{w(\bR)} = \ket{w(O_{3[111]} \bR)}, \nn \\
U_{C_{3[111]}}(\bk) \uef{} = \ket{\hat{u} (O_{C_{3[111]}}\bk)}.
\label{eq:cubic_SR}
\eg
This SR leads to a singular FT-CLS at $R=(\pi,\pi,\pi)$, which indicates a band crossing point of the FB at $R$.
As an example, we set $\ket{w(\bR)} = \ket{\bR,1}+\ket{\bR-\bb a_1,1}+\ket{\bR,2}+\ket{\bR-\bb a_2,2}+\ket{\bR,3}+\ket{\bR-\bb a_3,3}$ and $\uef{} = (1+\cm{Q_1},1+\cm{Q_2},1+\cm{Q_3})$.
The shape of the CLS is shown in \fig{Fig6}(b).
In fact, the band crossing at $R$ can also be explained by conventional representation theory.
At $R$, the symmetry operators satisfy
\bg
U_{C_{2x}}(R)^2 = U_{C_{2y}}(R)^2 = U_{C_{3[111]}}(R)^3 = \mathds{1}_3, \nn \\
U_{C_{3[111]}}(R) \, U_{C_{2x}}(R)=U_{C_{2y}}(R) \, U_{C_{3[111]}}(R), \nn \\
U_{C_{2x}}(R) \, U_{C_{3[111]}}(R) = U_{C_{3[111]}}(R) \, U_{C_{2y}}(R) \, U_{C_{2x}}(R), \nn \\
\left[ U_{C_{2x}}(R), U_{C_{2y}}(R) \right] = 0.
\eg
which are equivalent to the sufficient condition for having three-fold degeneracy as a three-dimensional irreducible representation of cubic symmetry group~\cite{bradlyn2016beyond}.

Now, we construct a KPC Hamiltonian,
\ba
H_{3D}(\bk)=\sum_{a=1}^3 \, \bsk{a}\bsb{a},
\ea
with three canonical BMOs, $\bsk{1}=(0,-1-Q_3,1+Q_2)$, $\bsk{2}=(1+Q_3,0,-1-Q_1)$, and $\bsk{3}=(-1-Q_2,1+Q_1,0)$.
The band structure is shown in \fig{Fig6}(c) where a FB has a band crossing point with three-fold degeneracy at $R$.

\section{Degenerate flat bands \label{sec:deg_flat}}
The KPC method can be extended to FB models with $n_F$-fold degenerate FB.
This can be done by using the BMOs whose rank is ${\rm max} \, N_B(\bk)=n_{\rm tot}-n_F$ when the total number of bands is $n_{\rm tot}$.
We note that this approach is best suited for the case when $n_F=n_{\rm tot}-1$ as only a single BMO can be used for the construction.
In this case, instead of the SR of the CLS, the SR of the BMO can be used to judge the presence of band crossing.
When there is no band crossing between the flat and dispersive bands, the degenerate FB can have either trivial or fragile topology~\cite{chiu2020fragile,peri2021fragile,cualuguaru2021general}.
We illustrate this point by constructing degenerate FBs in the Lieb lattice.
%

\begin{figure}[t!]
\centering
\includegraphics[width=0.45\textwidth]{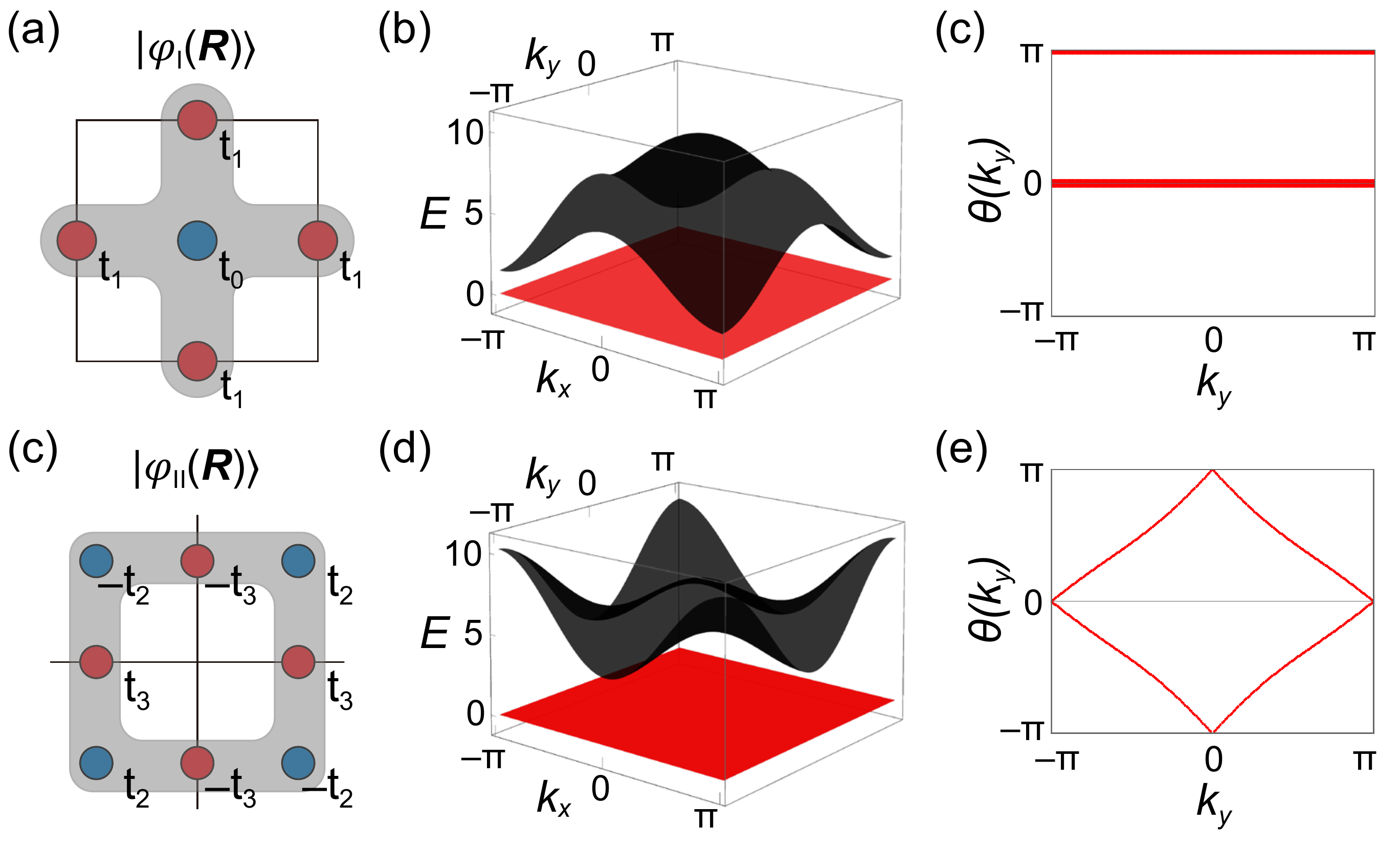}
\caption{
Degenerate FBs in the Lieb lattice.
(a)-(c) Description of FB model $H_{\rm deg,I}(\bk)$.
(a) The BMO $\bsk{\rm I}$.
(b) Band structure of $H_{\rm deg,I}(\bk)$ for $(t_0,t_1)=(1.2,1.0)$.
The FB (red) is two-fold degenerate.
(c) The Wilson loop spectrum of degenerate FB is trivial.
The Wilson loop spectrum $\theta(k_y)$ is calculated along the $k_x$ direction at each $k_y$.
(d)-(f) Description of FB model $H_{\rm deg,II}(\bk)$.
(d) The BMO $\bsk{\rm II}$.
(e) Band structure of $H_{\rm deg,II}(\bk)$ for $(t_2,t_3)=(0.8,1.0)$.
The FB (red) is two-fold degenerate.
(f) The Wilson loop spectrum of degenerate FB exhibits nontrivial winding indicating fragile band topology.
}
\label{Fig7}
\end{figure}

First, let us consider a single BMO with $\bb A_0$ SR under $C_4$: $U_{C_4}(\bk) \bsk{\rm I} = \ket{\vph_{\rm I}(O_{C_4}\bk)}$.
As an example, we choose $\bsk{\rm I}=\left( t_0,t_1(1+\cm{Q_1}),t_1(1+\cm{Q_2}) \right)$ whose real space representation is illustrated in \fig{Fig7}(a), and construct a KPC Hamiltonian,
\ba
H_{\rm deg,I}(\bk) = \bsk{\rm I} \bsb{\rm I}.
\ea
The band structure for $(t_0,t_1)=(1.2,1.0)$ is shown in \fig{Fig7}(b), where a FB with two-fold degeneracy exists without band crossing with the dispersive band.
The degenerate FB is topologically trivial, and this can be inferred in various ways.
For example, the BMO $\bskL{}{\rm I}{\bR}$ can be adiabatically tuned to the atomic orbital $\ket{\bR,1}$ by decreasing $t_1$ to zero.
Hence, the dispersive band has a band representation (BR), which is identical to the BR induced from $\ket{\bR,1}$.
The remaining bands, i.e. the degenerate FB, then have BR, identical to the BR corresponding to $\ket{\bR,2}$ and $\ket{\bR,3}$.
This proves the trivial band topology of FBs.
Also note that the Wilson loop spectrum (which is calculated for tight-binding Hamiltonian in the nonperiodic basis) does not exhibit any nontrivial winding, as shown in \fig{Fig7}(c).

Second, we now consider a single BMO $\bsk{\rm II}$ with $\bb D_2$ SR, which transforms as $U_{C_4}(\bk) \bsk{\rm II} = -Q_2 \ket{\vph_{\rm II}(O_{C_4}\bk)}$ in momentum space.
In \fig{Fig7}(d), one representative BMO, $\bsk{\rm II}=\left( t_2(1-Q_1)(1-Q_2),-t_3(1+Q_2),t_3(1+Q_1) \right)$, is shown.
Using this BMO, we construct a KPC Hamiltonian,
\ba
H_{\rm deg,II}(\bk) = \bsk{\rm II} \bsb{\rm II}.
\ea
The band structure for $(t_2,t_3)=(0.8,1.0)$ is shown in \fig{Fig7}(e), where a two-fold degenerate FB exists without band crossing with the dispersive band.
Contrary to the FBs in $H_{\rm deg,I}(\bk)$, the degenerate FB in $H_{\rm deg,II}(\bk)$ has fragile band topology~\cite{cano2018topology,bradlyn2019disconnected,bouhon2019wilson,else2019fragile,wieder2018axion,liu2019shift,bouhon2020geometric,alexandradinata2020crystallographic,peri2020experimental,zhang2021tunable}.
This is confirmed by obtaining the Wilson loop spectrum shown in \fig{Fig7}(f).
The Wilson loop spectrum exhibits nontrivial winding that indicates a fragile topology protected by $C_2$.
(This means that the fragile topology of FBs is intact even if we break $C_4$ but preserve $C_2$.)
Without calculating the Wilson loop spectrum, the fragile topology can be verified in other ways.
First, one can use symmetry indicators for fragile topology~\cite{hwang2019fragile,song2020fragile,song2020twisted} by inspecting symmetry eigenvalues of $C_2$ at high-symmetry points.
Second, the representation of FBs $\mc{R}_{\rm flat}$ can be expressed as a linear combination of BRs: In our case, $\mc{R}_{\rm flat} = \bb A_0 \oplus \bb B_+ \ominus \bb D_2$.
Here, $\bb A_l$ ($\bb D_l$) is a BR induced from a Wannier function whose Wyckoff position and $C_4$ eigenvalues are $\bb A$ ($\bb D$) and $e^{i\pi l/2}$, and $\bb B_\pm$ is a BR induced from a set of two Wannier functions whose Wyckoff position and $C_2$ eigenvalue are $\bb B$ and $\pm$.
The negative coefficient of $\bb D_2$ in $\mc{R}_{\rm flat}$ indicates fragile topology.

We have discussed the gapped degenerate FBs in the Lieb lattice as they are the most interesting cases.
However, degenerate FBs can also have band crossing points with other dispersive bands.
Such degenerate FB with a band crossing point can be constructed from a BMO with $\bb D_0$ SR, $\bsk{\rm III}=\left(1+Q_1+Q_2+Q_1 Q_2,t_4(1+Q_2),t_4(1+Q_1)\right)$ for example.
The $\bb D_0$ SR enforces a singular point of $\bsk{\rm III}$ at $M=(\pi,\pi)$, which corresponds to $(Q_1,Q_2)=(-1,-1)$.
Thus, any KPC Hamiltonian constructed by using $\bsk{\rm III}$ becomes the zero matrix at $M$ and the degenerate FB has a band crossing point at $M$.

\section{Summary and conclusion \label{sec:summary}}
In summary, we propose a general method for constructing FB models with and without band crossing points.
Using our method, a FB model which corresponds to a CLS with an arbitrary shape and SR can be straightforwardly constructed.
Thus, a fine tuning of hopping parameters on a case-by-case basis is unnecessary in our scheme for constructing FB models.
Especially, we exploit the fact that SFB has band crossing points while NSFB does not.
Also, as the singularity of FB is determined by the SR of CLS under unitary symmetry and its shape, our construction scheme naturally incorporates an important role of crystalline symmetries.
While, we have focused on nondegenerate FB mainly, we show that even degenerate FB can be obtained by using the KPC scheme with an example in the Lieb lattice.
Such degenerate FB can have both trivial and fragile band topology when the FBs are gapped from dispersive bands.
We anticipate that our construction scheme can be utilized to obtain various FB models in order to understand exotic topological and geometric properties of FB systems, which are appearing
in the fore of recent flat band studies.

Finally, we list potential applications and extensions of our work.
First, our construction scheme can be applied to realize nearly FBs with nontrivial topology.
Nearly FBs with nontrivial topology often arise when the degeneracy at the band crossing point of FB is lifted~\cite{wang2011nearly,bhattacharya2019flat,rhim2019classification,ma2020spin,hwang2021flat}.
Hence, once a FB model with a SFB is constructed, nearly FBs with nontrivial topology can be obtained by adding gap-opening perturbation which also breaks the exact flat dispersion of FB.
Our construction scheme can lead to an ideal tight-binding model as a basic platform for the study of exotic many-body phenomena such as the fractional Chern insulators~\cite{regnault2011fractional,andrews2018stability,andrews2021abelian}, which requires a nearly flat Chern band.
Second, known FB models can be generalized such that the hoppings beyond the nearest neighbor ones are included.
Usually, well-known FB models in some lattices such as Lieb and Kagome lattices involve only the nearest neighbor hoppings.
Once the relevant BMOs are found, the extension of FB models can be straightforwardly achieved through the KPC scheme.
Finally, we comment that, for $n_F$-fold degenerate FB, choosing the relevant BMOs is complicated unless $n_F=n_{\rm tot}-1$ which corresponds to the case discussed in Sec.~\ref{sec:deg_flat}.
It is because the BMOs must be orthogonal to all the FT-CLSs while our prescriptions proposed in Sec.~\ref{subsec:KPC_basis} are best applied to nondegenerate FB.
Thus, extending our work in a more practical way to nonsymmorphic lattices, where degenerate FB must arise, is an important future work.

{\it Note added.|}
Recently, we became aware of \Rf{graf2021designing} by A. Graf and F. Pi{\'e}chon where FB models with multi-fold band crossing points are systematically constructed by using the canonical BMOs.

\begin{acknowledgments}
Y.H. and B.-J.Y. were supported by the Institute for Basic Science in Korea (Grant No. IBS-R009-D1), 
Samsung  Science and Technology Foundation under Project Number SSTF-BA2002-06,
the National Research Foundation of Korea (NRF) Grant funded by the Korea government (MSIT) (No. 2021R1A2C4002773, and No. NRF-2021R1A5A1032996).
J.-W.R. was supported by Institute for Basic Science in Korea (Grant No. IBSR009-D1), the National Research Foundation of Korea (NRF) Grant funded by the Korea government (MSIT) (Grant No. 2021R1A2C101057211).
\end{acknowledgments}

\appendix
\section{Tight-binding Hamiltonian in periodic basis \label{app:TB}}
In general, a tight-binding Hamiltonian $\hat{H}$ is expressed as
\ba
\hat{H} = \sum_{\bR,\Delta \bR} \sum_{\alpha,\beta} t_{\alpha \leftarrow \beta}(\Delta \bR) \ket{\bR+\Delta \bR, \alpha} \bra{\bR,\beta},
\ea
with hopping parameters $t_{\alpha \leftarrow \beta}(\Delta \bR)$ ($\alpha,\beta=1,\dots,n_{\rm tot}$).
A Fourier transform of atomic orbitals $\ket{\bR,\alpha}$ reduces $\hat{H}$ to $n_{\rm tot} \times n_{\rm tot}$ Hamiltonian in momentum space.
There are two standard representations for the tight-binding Hamiltonian in momentum space, obtained by the periodic and nonperiodic bases, respectively.
These two bases are given by two different ways of performing Fourier transforms,
\ba
\label{eq:basis_def}
&\ket{\bk,\alpha}
\equiv \frac{1}{\sqrt{N_{\rm cell}}} \sum_{\bR} e^{i \bk \cdot \bR} \, \ket{\bR,\alpha}, \\
&\ket{\bk,\alpha}'
\equiv \frac{1}{\sqrt{N_{\rm cell}}} \sum_{\bR} e^{i \bk \cdot (\bR + \bx_\alpha)} \, \ket{\bR,\alpha},
\ea
respectively, where $N_{\rm cell}$ is the number of unit cells in the periodic lattice system.
Note that inverse Fourier transforms,
\ba
\ket{\bR, \alpha} &= \frac{1}{\sqrt{N_{\rm cell}}} \sum_{\bk} e^{-i \bk \cdot \bR} \, \ket{\bk,\alpha}, \\
&= \frac{1}{\sqrt{N_{\rm cell}}} \sum_{\bk} e^{-i \bk \cdot (\bR+\bx_\alpha)} \, \ket{\bk,\alpha}',
\ea
can be performed by using $\sum_{\bR} e^{i (\bk-\bk') \cdot \bR} = N_{\rm cell} \delta_{\bk,\bk'}$ and $\sum_{\bk} e^{i \bk \cdot (\bR-\bR')} = N_{\rm cell} \delta_{\bR,\bR'}$.
The tight-binding Hamiltonian in the periodic basis is expressed as
\ba
H(\bk)_{\alpha\beta} = \sum_{\bR} \, t_{\alpha \leftarrow \beta}(\bR) \, e^{-i \bk \cdot \bR},
\ea
while the other one in the nonperiodic basis is given by
\ba
H'(\bk)_{\alpha\beta} = \sum_{\bR} \, t_{\alpha \leftarrow \beta}(\bR) \, e^{-i \bk \cdot (\bR+\bx_\alpha-\bx_\beta)}.
\ea
The periodic and nonperiodic bases can be changed by using the sublattice embedding matrix $V(\bk)_{\alpha \beta} = e^{-i \bk \cdot \bx_\alpha} \delta_{\alpha \beta}$: $H(\bk) = V(\bk)^{-1} H'(\bk) V(\bk)$.
The physical observables must be obtained in the nonperiodic basis, because the information on the sublattice sites $\bx_\alpha$ are omitted in the periodic basis.
Nevertheless, the periodic basis is useful not only for topological classification~\cite{shiozaki2017topological,read2017compactly} but also for constructing FB models due to its periodicity in the BZ.
Now, let us consider a unitary symmetry $\hat{\sg}=\{O_\sg|\bb \delta_\sg\}$ which acts on real-space coordinates $\bb r$ as $\hat{\sg}: \bb R \rightarrow O_\sg \bb R + \bb \delta_\sg$.
$\hat{\sg}$ also acts on the atomic orbitals such that
\ba
\hat{\sg} \, \ket{\bR,\alpha}
&= \ket{\bR_\sg(\alpha),\beta} \, U(\sg)_{\beta \alpha},
\label{eq:unisym_real}
\ea
where $\bR_\sg(\alpha)= O_\sg \bR + O_\sg \bx_\alpha + \bb \delta_\sg - \bx_\beta$.
Then, symmetry operator $U_\sg(\bk)$ is defined by a symmetry transformation of $\ket{\bk,\alpha}$ under $\hat{\sg}$:
\ba
\hat{\sg} \, \ket{\bk,\alpha}
= \ket{O_\sg \bk,\beta} \, U_\sg(\bk)_{\beta \alpha}.
\label{eq:unisym_BZ}
\ea
Explicitly, $U_\sg(\bk)$ is given by 
\ba
U_\sg(\bk)
&= V(O_\sg \bk)^\dg U(\sg) V(\bk) \\
&= e^{i O_\sg \bk \cdot (\bx_\beta - O_\sg \bx_\alpha -\bb \delta_\sg)} \, U(\sg).
\label{eq:Uk_def}
\ea
The symmetry operator $U_\sg(\bk)$ defines a symmetry transformation of $H(\bk)$, $H(O_\sg \bk) = U_\sg (\bk) \, H(\bk) \, U_\sg(\bk)^\dg$.

Finally, we summarize two useful formulas.
First, for $\hat{\sg}=\hat{\sg}_2 \hat{\sg}_1$, symmetry operator of $\sg$ is given by $U_{\sg_2 \sg_1}(\bk) = U_{\sg_2}(O_{\sg_1}\bk) U_{\sg_1}(\bk)$.
Second, $q$-fold symmetry $\hat{\sg}$ satisfies
\ba
U_\sg(O_\sg^{q-1}\bk) U_\sg(O_\sg^{q-2}\bk) \cdots U_\sg(\bk) = \pm e^{-i \bk \cdot \bb \Delta_\sg},
\label{eq:unisym_fold}
\ea
with a lattice vector $\Delta_\sg$.
The lattice vector $\Delta_\sg$ is determined by $\hat{\sg}^q = \pm \hat{t}(\bb \Delta_\sg)$ where $\hat{t}(\bb \Delta_\sg)$ denote a translation by $\bb \Delta_\sg$.

\section{Symmetrization algorithm \label{app:symmetrization}}
Consider a CLS having a definite symmetry representation with respect to symmetry group $G$.
Then, as discussed in Sec.~\ref{subsec:KPC_symmetrization}, the FT-CLS transforms as
\ba
U_\sg(\bk) \uef{} = \ket{\hat{u}(O_\sg \bk)} \, \symeigv(\bk),
\label{eq:symrepsg}
\ea
with respect to $\sg \in G$.
Once we obtain a FB model through the KPC, the resulting tight-binding Hamiltonian $H_0(\bk)_{\alpha \beta}=\bra{\bk,\alpha}\hat{H}_0\ket{\bk,\beta}$ may not be symmetric under $G$.

Now, we review the symmetrization algorithm for the tight-binding Hamiltonian~\cite{gresch2018automated}.
Before discussing the general case, let us first consider $G=p6$ for spinless electron as an example.
$G$ is generated by $C_6$ rotation, thus $\sg \in \{C_6^p|p=0,1,\dots,5\}$.
For the symmetrization with respect to $C_6$, we define $\hat{H}_{\rm sym}$:
\ba
\hat{H}_{\rm sym} = \sum_{p=0}^{5} \hat{C}_6^{-p} \hat{H}_0 \hat{C}_6^p,
\ea
then, $\hat{\sg} \hat{H}_{\rm sym} \hat{\sg}^{-1} = \hat{H}_{\rm sym}$ is automatically satisfied, since $\hat{C}_6^6$ is equal to a  translation operator $\hat{t}(\bb \Delta)$ for a certain lattice vector $\bb \Delta$.
The resulting Hamiltonian $\hat{H}_{\rm sym}$ in the basis $\ket{\bk,\alpha}$, $H_{\rm sym}(\bk)_{\alpha \beta}=\bra{\bk,\alpha} \hat{H}_{\rm sym} \ket{\bk,\beta}$, is given by
\ba
H_{\rm sym}(\bk)_{\alpha \beta} = \sum_{p=0}^{5} \, H^{(p)}_0(\bk)_{\alpha \beta},
\ea
where we define $H^{(p)}_0(\bk)_{\alpha \beta}$ as
\ba
H^{(p)}_0(\bk)_{\alpha \beta}
&=\bra{\bk,\alpha} \hat{C}_6^{-p} \hat{H}_0 \hat{C}_6^p \ket{\bk,\beta} \nn \\
&=\sum_{p=0}^5 \, U_{C_6^p}(\bk)^\dg H_0(O_{C_6^p}\bk) U_{C_6^p}(\bk).
\ea

The above example clearly shows that a symmetrization of Hamiltonian can be done by adding symmetry image of the original Hamiltonian with respect to all the symmetry elements in $G$.
Thus, this result is generalized to a general $G$~\cite{gresch2018automated}:
\bg
\hat{H}_{\rm sym} = \sum_{\sg \in G} \, \hat{\sg}^{-1} \hat{H}_0 \hat{\sg}, \nn \\
H_{\rm sym}(\bk) = \sum_{\sg \in G} \, U_\sg(\bk)^\dg H_0(O_\sg \bk) U_\sg(\bk).
\label{eq:symalgorithm}
\eg
During this procedure, only symmetry-allowed hoppings survive, and one obtains the tight-binding Hamiltonian that is symmetric under the symmetry group $G$.
Also, we comment that the symmetrization algorithm can also be applied to antiunitary symmetry by combining unitary symmetry and the complex conjugation.
%

\begin{figure}[t!]
\centering
\includegraphics[width=0.42\textwidth]{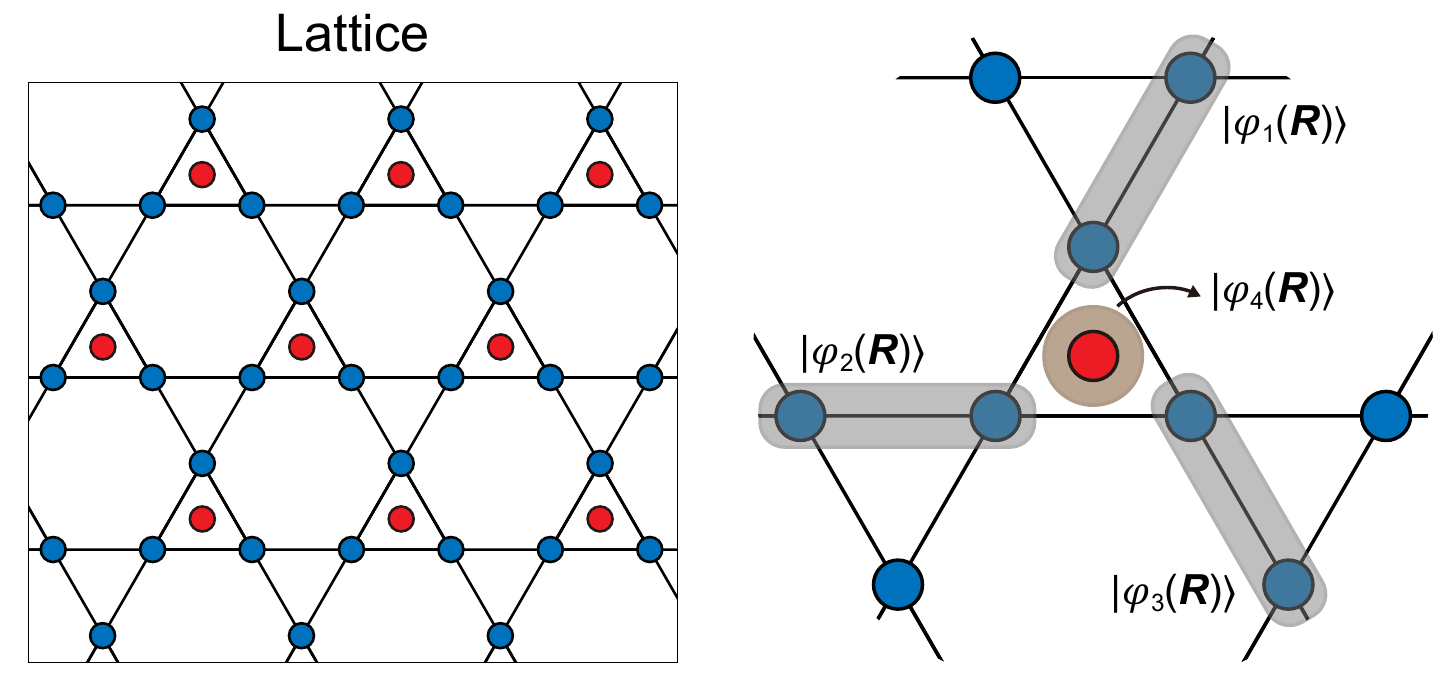}
\caption{BMOs in $C_3$ symmetric lattice.
The $C_3$-rotation axis coincides with the red sublattice.
The real-space representations of BMOs $\ket{\vph_a(\bR)}$ are illustrated schematically in gray and brown regions.
}
\label{FigS1}
\end{figure}

Now, we prove that the FB remains even after the symmetrization algorithm is implemented for KPC Hamiltonian $H_0(\bk)$: $H_0(\bk)=\sum_{a,b=1,\dots,n_B}f_{ab}(\bk) \kbr{\vph_a(\bk)}{\vph_b(\bk)}$ with BMOs $\bsk{a}$ satisfying $\brk{\vph_a(\bk)}{\hat{u}(\bk)}=0$.
From \eqs{eq:symrepsg} and \eqref{eq:symalgorithm}, we show that
\ba
H_{\rm sym}(\bk) \uef{} =& \sum_{\sg \in G} \, U_\sg(\bk)^\dg H_0(O_\sg \bk) U_\sg(\bk) \uef{} \nn \\
=& \sum_{\sg \in G} \, U_\sg(\bk)^\dg \, \left[ H_0(O_\sg \bk) \ket{\hat{u}(O_\sg \bk)} \right] \, \symeigv(\bk) \nn \\
=&0,
\ea
and hence $\uef{}$ is still the FT-CLS of $H_{\rm sym}(\bk)$.
This result can also be generalized to degenerate FB.
In general, CLSs $\ket{w_{A}(\bR)}$ ($A=1,\dots,n_F$) of $n_F$-fold degenerate FB can be chosen such that each CLS is symmetric alone or mapped to another CLS under $\sg$.
Hence, the FT-CLSs $\uef{A}$ transform under $\sg$ as
\ba
U_\sg(\bk) \uef{A} = \uef{A'} B_\sg(\bk)_{A'A},
\ea
where $B_\sg(\bk)_{A'A}$ denotes the sewing matrix element between $\uef{A}$ and $\uef{A'}$.
Hence, $H_{\rm sym}(\bk) \uef{A} \propto H_0(O_\sg \bk) \ket{\hat{u}_{A'}(O_\sg \bk)} \, B_\sg(\bk)_{A'A}=0$ when the BMOs satisfy $\brk{\vph_a(\bk)}{\hat{u}_A(\bk)}=0$ for $a=1,\dots,n_B$ and $A=1,\dots,n_F$.

Finally, we comment on the BMOs $\bsk{a}$.
If these BMOs are permuted by $G$ or each BMO transforms as SR under $G$, then the BMOs respect $G$.
For example, we list the BMOs $\bsk{a}$ ($a=1,\dots,4$) that respect $C_3$ rotation:
\ba
U_{C_3}(\bk) \bsk{1} &= \bs{2}{O_{C_3}\bk}, \nn \\
U_{C_3}(\bk) \bsk{2} &= \bs{3}{O_{C_3}\bk}, \nn \\
U_{C_3}(\bk) \bsk{3} &= \bs{1}{O_{C_3}\bk}, \nn \\
U_{C_3}(\bk) \bsk{4} &= \bs{4}{O_{C_3}\bk} e^{i\frac{2\pi}{3}}.
\label{eq:C3SR}
\ea
In terms of real-space representation $\bs{a}{\bR}$ corresponding to $\bsk{a}$, \eq{eq:C3SR} is equivalent to
\ba
\hat{C_3} \bs{1}{\bR} &= \bs{2}{O_{C_3}\bR}, \nn \\
\hat{C_3} \bs{2}{\bR} &= \bs{3}{O_{C_3}\bR}, \nn \\
\hat{C_3} \bs{3}{\bR} &= \bs{1}{O_{C_3}\bR}, \nn \\
\hat{C_3} \bs{4}{\bR} &= \bs{4}{O_{C_3}\bR} e^{i\frac{2\pi}{3}},
\ea
as shown in \fig{FigS1}.
However, some BMOs $\bskL{}{\rm old}{\bk}$ may not satisfy the above condition for respecting $G$.
In this case, the symmetrization algorithm generates additional BMOs $\bskL{}{\rm new}{\bk}$ so that $\bskL{}{\rm old}{\bk}$ and $\bskL{}{\rm new}{\bk}$ are permuted under $G$.
For example, let us construct the KPC Hamiltonian using only $\bsk{1}$ in \eq{eq:C3SR}, $H_{\rm KPC}(\bk)=f_{11}(\bk) \bsk{1}\bsb{1}$.
After the symmetrization, we have
\ba
H_{\rm KPC, sym}(\bk) &= f_{11}(\bk) \bsk{1}\bsb{1} \nn \\
&+ f_{22}(O_{C_3}^{-1}\bk) \bsk{2}\bsb{2} \nn \\
&+ f_{33}(O_{C_3}\bk) \bsk{3}\bsb{3},
\ea
which is symmetric under $C_3$.

\section{Flat-band models protected by \texorpdfstring{$C \circ I_{ST}$}{CI_{ST}} symmetry \label{app:exceptional}}
As mentioned in Sec.~\ref{subsec:NSFB_Lieb}, the spin-orbit coupled Lieb model~\cite{weeks2010topological} cannot be constructed through the KPC scheme.
The tight-binding Hamiltonian and the FT-CLS of the spin-orbit coupled Lieb are given by
\bg
H''_{\rm Lieb}(\bk) = t_0 \bpm
0 & (1+Q_1) & (1+Q_2) \\
(1+\cm{Q_1}) & 0 & g_{\rm soc}(\bk) \\
(1+\cm{Q_2}) & \cm{g_{\rm soc}(\bk)} & 0
\epm, \\
\uefL{''}{\rm Lieb} = \bpm i \lambda (1-Q_1)(1-Q_2) \\ -t_0-t_0 Q_2 \\ t_0+t_0 Q_1 \epm,
\eg
respectively, where $g_{\rm soc}(\bk)=i\frac{\lambda}{t_0}(1-\cm{Q_1})(1-Q_2)$.
For a given FT-CLS $\uefL{''}{\rm Lieb}$, let us set three BMOs canonically:
\ba
\bskL{''}{1}{\bk} &= \left( 0,1+\cm{Q_1},1+\cm{Q_2} \right), \nn \\ \bskL{''}{2}{\bk} &= \left( 1+Q_1,0,-i\frac{\lambda}{t_0} (1-Q_1)(1-\cm{Q_2}) \right), \nn \\
\bskL{''}{3}{\bk} &= \left( 1+Q_2, i\frac{\lambda}{t_0}(1-\cm{Q_1})(1-Q_2), 0 \right).
\ea
Although $H''_{\rm Lieb}(\bk)$ can be expressed as the sum of Kronecker products of BMOs as
\bg
H''_{\rm Lieb}(\bk) = f''_{12}(\bk) h''_{12}(\bk) + \cm{f''_{12}(\bk)} h''_{21}(\bk),
\eg
where $f''_{12}(\bk)=-t_0(1+Q_1)^{-1}$ and $h''_{ab}(\bk)=\bskL{''}{a}{\bk}\bsbL{''}{b}{\bk}$ ($a,b=1,2$), our assumption that $f''_{ab}(\bk)$ should be a Laurent polynomial is violated.

The existence of FB in this model can be explained by $C \circ I_{ST}$, antiunitary symmetry combined with chiral $C$ and space-time inversion $I_{ST}$ symmetries.
In the presence of $C \circ I_{ST}$, symmetry constraint on the Hamiltonian $H(\bk)$ is given by
\ba
U_{C \circ I_{ST}}(\bk) \, \cm{H(\bk)} \, U_{C \circ I_{ST}}(\bk)^{-1} = -H(\bk).
\label{eq:CIst}
\ea
Note that \eq{eq:CIst} implies $U_{C \circ I_{ST}}(\bk) \cm{U_{C \circ I_{ST}}(\bk)} = \mathds{1}$.
Hence, $C \circ I_{ST}$ gives rise to a symmetric band structure, $\{E(\bk)\}=\{-E(\bk)\}$, in the sense that $-E(\bk)$ is also an energy eigenvalue at $\bk$ when $E(\bk)$ is an energy eigenvalue at $\bk$.
Accordingly, one of the bands must be flat when the number of bands is an odd integer.
It is worth comparing $C \circ I_{ST}$ with chiral symmetry $C$.
Chiral symmetry $C$ constrains the band structure in the same way as $C \circ I_{ST}$: $\{E(\bk)\}=\{-E(\bk)\}$.
However, when $C$ satisfies ${\rm Tr}[U_C]=\pm n_F$, there are $n_F$ number of FBs at zero energy~\cite{lieb1989two}.
In contrast, $C \circ I_{ST}$ protects a single FB only when the total number of bands is an odd integer.
%

\begin{figure*}[t!]
\centering
\includegraphics[width=0.9\textwidth]{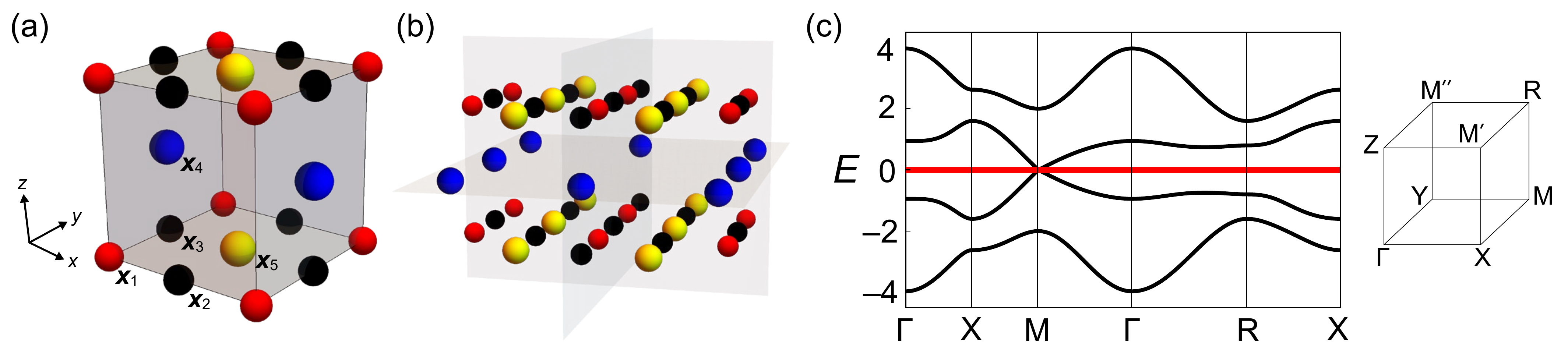}
\caption{
Flat band in 3D five-band system.
(a) Sublattices are located at $\bx_1=(0,0,0)$, $\bx_2=(1/2,0,0)$, $\bx_3=(0,1/2,0)$, $\bx_4=(0,1/2,1/2)$, and $\bx_5=(1/2,1/2,0)$, respectively.
(b) Schematic illustration of the CLS.
We illustrate only the shape of CLS.
(c) Band structure of $\widetilde{H}_{\rm five}(\bk)$ for $(t_0,t_1,t_2,\lambda_0,\lambda_1,\lambda_2,\lambda_3)=(1.0,0.8,0.6,0.4,0.3,0.2,0.2)$.
A FB is denoted by the red line.}
\label{FigS2}
\end{figure*}

From now on, we consider a tight-binding Hamiltonian $\widetilde{H}(\bk)$ in the nonperiodic basis.
In this basis, one can find a coordinate system where symmetry operator for $C \circ I_{ST}$ is independent of $\bk$: $\widetilde{U}_{C \circ I_{ST}}(\bk)=U(C \circ I_{ST})$.
We first consider the three-band system to which the spin-orbit coupled Lieb model belongs.
In the presence of $C \circ I_{ST}$ with $U(C \circ I_{ST})={\rm Diag}(-1,1,1)$, symmetry constraint in \eq{eq:CIst} becomes
\ba
\widetilde{H}(\bk)=
\bpm
0 & M_{12}(\bk) & M_{13}(\bk) \\
M_{12}(\bk) & 0 & iM_{23}(\bk) \\
M_{13}(\bk) & -iM_{23}(\bk) & 0
\epm,
\ea
where $[M_{12}(\bk),M_{13}(\bk),M_{23}(\bk)]$ are real functions of $\bk$.
Also, energy eigenvalues are given by 0 and $\pm [M_{12}(\bk)^2+M_{13}(\bk)^2+M_{23}(\bk)^2]^{1/2}$.

For the spin-orbit coupled Lieb model, we obtain a tight-binding Hamiltonian in the nonperiodic basis using the sublattice embedding matrix $V(\bk)={\rm Diag}(1,e^{-i k_x/2},e^{-i k_y/2})$:
\ba
\widetilde{H}''_{\rm Lieb}(\bk)
&= V(\bk) H''_{\rm Lieb}(\bk) V(\bk)^{-1} \nn \\
&= \bpm 0 & M_{12}(\bk) & M_{13}(\bk) \\
M_{12}(\bk) & 0 & iM_{23}(\bk) \\
M_{13}(\bk) & -iM_{23}(\bk) & 0
\epm,
\ea
where $M_{12}(\bk)=2t_0 \cos \frac{k_x}{2}$, $M_{13}(\bk)=2t_0 \cos \frac{k_y}{2}$, and $M_{13}(\bk)=4\lambda \sin \frac{k_x}{2} \sin \frac{k_y}{2}$.
Hence, $\widetilde{H}''_{\rm Lieb}(\bk)$ has the same form of $\widetilde{H}(\bk)$ and the FB at zero energy is protected by $C \circ I_{ST}$.

For five-band system, we consider $U(C \circ I_{ST})={\rm Diag}(-1,1,1,-1,-1)$.
Then, symmetry constraint in \eq{eq:CIst} becomes
\ba
\widetilde{H}(\bk)=
\bpm
0 & M_{12} & M_{13} & iM_{14} & iM_{15} \\
M_{12} & 0 & iM_{23} & M_{24} & M_{25} \\
M_{13} & -iM_{23} & 0 & M_{34} & M_{35} \\
-iM_{14} & M_{24} & M_{34} & 0 & iM_{45} \\
-iM_{15} & M_{25} & M_{35} & -iM_{45} & 0 \epm,
\ea
where $M_{\alpha\beta}$ ($\alpha,\beta=1,\dots,5$) is a real function of $\bk$.
It is crucial to note that every element of tight-binding Hamiltonian in the periodic basis corresponding to $\widetilde{H}(\bk)$ must be a Laurent polynomial in variables $Q_i$ ($i=1,\dots,d)$.
Otherwise, $\widetilde{H}(\bk)$ cannot be realized with finite-range hoppings.
For this, we consider a lattice system in 3D ($d=3$) where the sublattices are located at $\bx_1=(0,0,0)$, $\bx_2=(1/2,0,0)$, $\bx_3=(0,1/2,0)$, $\bx_4=(0,1/2,1/2)$, and $\bx_5=(1/2,1/2,0)$, as shown in \fig{FigS2}(a).
\begin{widetext}
In this lattice system, we construct a FB model $\widetilde{H}_{\rm five}(\bk)$:
\ba
\widetilde{H}_{\rm five}(\bk) = \bpm
0 & 2t_0 \cos \frac{k_x}{2} & 2t_0 \cos \frac{k_y}{2} & 4i\lambda_1 \cos \frac{k_y}{2} \cos \frac{k_z}{2} & 4i\lambda_2 \cos \frac{k_x}{2} \cos \frac{k_y}{2} \\
2t_0 \cos \frac{k_x}{2} & 0 & 4i\lambda_0 \sin \frac{k_x}{2} \sin \frac{k_y}{2} & 0 & 2t_1 \cos \frac{k_y}{2} \\
2t_0 \cos \frac{k_y}{2} & -4i\lambda_0 \sin \frac{k_x}{2} \sin \frac{k_y}{2} & 0 & 2t_2 \cos \frac{k_z}{2} & 2t_1 \cos \frac{k_x}{2} \\
-4i\lambda_1 \cos \frac{k_y}{2} \cos \frac{k_z}{2} & 0 & 2t_2 \cos \frac{k_z}{2} & 0 & 4i\lambda_3 \sin \frac{k_z}{2} \sin \frac{k_x}{2} \\
-4i\lambda_2 \cos \frac{k_x}{2} \cos \frac{k_y}{2} & 2t_1 \cos \frac{k_y}{2} & 2t_1 \cos \frac{k_x}{2} & -4i\lambda_3 \sin \frac{k_z}{2} \sin \frac{k_x}{2} & 0 \epm.
\ea
\end{widetext}
Note that hoppings $t_{\alpha \leftarrow \beta}(\Delta \bR)$ can be read off from the Hamiltonian in periodic basis $H_{\rm five}(\bk)=V(\bk)^{-1} \widetilde{H}_{\rm five}(\bk) V(\bk)$ by using \eq{eq:hopping}.
In \figs{FigS2}(b) and \hyperref[FigS2]{9}(c), the CLS and band structure are shown.

%

\end{document}